\documentclass[pre,twocolumn,amsmath,amssymb,floatfix,superscriptaddress,eqsecnum]{revtex4}
\usepackage{hyperref}
\usepackage[utf8x]{inputenc}
\usepackage{bm}
\usepackage{color}
\usepackage{graphicx,epsfig}
\newcommand{\tb}{\mathring{\tau}}

\newcommand{\cb}{\mathring{c}}

\newcommand{\rmd}{\mathrm{d}}
\newcommand{\rme}{\mathrm{e}}
\newcommand{\rmi}{\mathrm{i}}
\newcommand{\pint}{\int_{\bm{p}}^{(d-1)}}
\newcommand{\Tr}{\mathop{\mathrm{tr}}}
\newcommand{\fex}{f_{\mathrm{ex}}}
\newcommand{\fb}{f_{\mathrm{b}}}
\newcommand{\fs}{f_{\mathrm{s}}}
\newcommand{\arsinh}{\mathop{\mathrm{arsinh}}}
\newcommand{\Leff}{L_{\mathrm{eff}}}
\newcommand{\Thc}{c}
\newcommand{\Thd}{d}
\newcommand{\mat}[1]{\mathbf{#1}}
\newcommand{\D}[1]{\Delta_{\mathrm{C}#1}}
\newcommand{\aseq}{\simeq}
\newcommand{\asprop}{\sim}
\newcommand{\etwa}{\approx}

\newcommand{\ellipK}{K}

\begin{document}

\title{Large-$n$ approach to  thermodynamic Casimir effects in slabs with free surfaces} 
\author{H.~W. Diehl}
\affiliation{Fakultät für Physik, Universität Duisburg-Essen, D-47048 Duisburg, Germany}

\author{Daniel Grüneberg}
\affiliation{Fakultät für Physik, Universität Duisburg-Essen, D-47048 Duisburg, Germany}

\author{Martin Hasenbusch}
\affiliation{Institut für Physik, Humboldt-Universität zu Berlin, Newtonstr.~15, D-12489 Berlin, Germany}

\author{Alfred Hucht}
\affiliation{Fakultät für Physik, Universität Duisburg-Essen, D-47048 Duisburg, Germany}

\author{Sergei B. Rutkevich}
\affiliation{Fakultät für Physik, Universität Duisburg-Essen, D-47048 Duisburg, Germany}
\affiliation{on leave from Institute of Solid State and Semiconductor Physics, Minsk, Belarus}

\author{Felix M. Schmidt}
\affiliation{Fakultät für Physik, Universität Duisburg-Essen, D-47048 Duisburg, Germany}

\date{\today}

\begin{abstract}
The classical $n$-vector $\phi^4$ model with $O(n)$ symmetrical Hamiltonian $\mathcal{H}$ is considered in a $\infty^2\times L$ slab geometry bounded by a pair of parallel free surface planes at separation $L$. Standard quadratic boundary terms implying Robin boundary conditions are included in $\mathcal{H}$. The temperature-dependent scaling functions of the excess free energy and the thermodynamic Casimir force are computed in the large-$n$ limit for temperatures $T$  at, above, and below the bulk critical temperature $T_{\mathrm{c}}$. Their  $n=\infty$ limits can be expressed exactly in terms of the spectrum and eigenfunctions of a  self-consistent one-dimensional Schrödinger equation. This equation is solved by numerical means for two distinct discretized versions of the model: in the first (``model A''), only the coordinate $z$ across the slab is discretized and the integrations over momenta conjugate to the lateral coordinates are regularized dimensionally; in the second (``model B''), a simple cubic lattice with periodic boundary conditions along the lateral directions is used. Renormalization-group ideas are invoked to show that, in addition to corrections to scaling  $\propto L^{-1}$, anomalous ones $\propto L^{-1}\ln L$ should occur. They can be  considerably decreased  by taking an appropriate $g\to\infty $ ($T_{\mathrm{c}}\to\infty$) limit of the  $\phi^4$ interaction constant $g$. Depending on the model A or B, they can be  absorbed completely or to a large extent in an effective thickness $L_{\mathrm{eff}}=L+\delta L$. Excellent data collapses and consistent high-precision results for both models are obtained. The approach to the low-temperature Goldstone values of the scaling functions is shown to involve logarithmic anomalies. The scaling functions exhibit all qualitative features seen in experiments on the thinning of wetting layers of ${}^4$He and Monte Carlo simulations of $XY$ models, including a pronounced minimum of the Casimir force below $T_{\mathrm{c}}$. The results are in conformity with various analytically known exact properties of the scaling functions.
 \end{abstract}

\maketitle
\section{Introduction}

The confinement of low-energy long-wavelength fluctuations in media frequently leads to effective forces between confining boundaries and macroscopic bodies that are immersed into these media. A celebrated and much studied class of examples of such fluctuation-induced forces are the quantum electrodynamics (QED) Casimir forces that act between arbitrary objects coupling to the electromagnetic field, such as grounded metallic plates in vacuum. They are caused by the confinement of vacuum fluctuations of the electromagnetic field \cite{Cas48,casiqmrev,BMM01}.

During the past two decades, it has become clear that a wealth of classical analogs of such effective forces, called ``thermodynamic Casimir forces,'' exist \cite{casitdrev,Kre94,Kre99,BDT00}. Rather than by quantum fluctuations, the latter are induced by thermal fluctuations at or near critical and multicritical points, or by Goldstone modes \cite{KG99}. The purpose of the present paper is to present exact results for the thermodynamic Casimir force of the $O(n)$ $\phi^4$ model in an $\infty^2\times L$ slab geometry bounded by two free surfaces at $z=0$ and $z=L$. A brief account of some of the results reported here was given in a recent letter \cite{DGHHRS12}. The issue has also been taken up in a recent work \cite{BDR12xxx,DBR14}, where parts of the results of \cite{DGHHRS12} were reproduced (to a considerably lower accuracy) \cite{rem:BDR}.

Our motivation for this work is rooted in the following considerations. The universal properties of QED Casimir forces at temperature $T=0$ usually can be studied within the framework of effective free field theories in which the interaction of the electromagnetic fields with matter is taken into account via appropriate boundary conditions at the surfaces of the confining material bodies. By contrast, for adequate investigations of thermodynamic Casimir forces near critical and multicritical points, the use of interacting field theories is indispensable. Studies of such kind, though important and of general interest, normally turn out to be rather challenging because they involve a combination of difficult problems. Satisfactory theories of thermodynamic Casimir forces in $d$-dimensional slabs as functions of temperature and thickness $L$ must be able to cope with bulk and boundary critical behavior, finite-size critical or pseudocritical behavior, and the crossover from $d$- to $(d{-}1)$-dimensional behavior that occurs as the  bulk correlation length $\xi$ becomes larger than $L$. 
Furthermore, they should yield a qualitatively correct phase diagram for finite $L$.
This demands a proper treatment of low-temperature excitations.

In order to safeguard against misunderstandings, a few remarks are appropriate. Note, first of all, that we shall not consider temperature effects on QED Casimir forces. One source of such effects is thermal fluctuations in the material bodies immersed into the QED vacuum. If one chooses instead of the QED vacuum a polarizable and magnetizable medium, a second source of temperature effects is thermal fluctuations in the medium. Both types of temperature effects have attracted considerable attention and occasionally produced controversial results \cite{KMM09}. They depend on properties of the models of matter chosen for the material bodies and the medium, and they exhibit a lesser degree of universality than $T=0$ QED Casimir forces and thermodynamic Casimir forces. We shall not engage in their analysis in this paper. Second, following  established practice, we will refer to effective forces induced by thermal fluctuations near critical (or multicritical) points as thermodynamic Casimir forces, using the adjective ``thermodynamic'' to indicate that temperatures other than $T_{\mathrm{c}}$ are also considered. 

Instructive examples of the kind of systems we will be concerned with are provided by ${d=3}$~dimensional systems whose low-temperature bulk phase exhibits long-range order and the spontaneous breaking of a continuous symmetry. To become specific, take a classical $n$-vector $\phi^4$ model on a slab $\mathbb{R}^2\times[0,L]$ whose Hamiltonian $\mathcal{H}[\bm{\phi}]$  is invariant under the symmetry group $O(n)$ and involves only short-range interactions. In the thermodynamic bulk limit $L=\infty$, a bulk critical temperature $T_{\mathrm{c}}>0$ exists above and below which the model is disordered and ordered, respectively. The spontaneous breaking of the symmetry from $O(n)$ to   $O(n-1)$ in the low-temperature phase implies the presence of Goldstone modes (``spin waves''). For finite $L$, low-energy excitations of this kind destroy long-range order for any $T>0$: it is rigorously known  that no phase with long-range order exists at $T>0$ when $L<\infty$; see, e.g., \cite{MW66,JF71,MW94}. Thus, a crossover from three-dimensional bulk critical behavior to two-dimensional pseudocritical behavior should occur as the bulk correlation length $\xi$ becomes larger than $L$. This applies to the disordered phase as $T\to T_{\mathrm{c}}$. In the ordered bulk phase, $\xi=\infty$ because of Goldstone singularities at any $T<T_{\mathrm{c}}$; then the appropriate length scale $\lesssim L$ up to which bulk behavior locally prevails  is given by the Josephson coherence length \cite{Jos66} (spin stiffness; see, e.g., \cite{CL95}). Furthermore, the reduced thermodynamic Casimir force per unit area, $\beta \mathcal{F}_{\mathrm{C}}(T,L)$, where $\beta=1/k_{\mathrm{B}}T$, does not vanish in the limit $T\to 0$ since confined Goldstone-mode excitations give rise to a fluctuation-induced force \cite{KG99}.

The features just discussed manifest themselves in the temperature dependence of the thermodynamic Casimir force $\beta \mathcal{F}_\mathrm{C}(T,L)$. Recall that according to finite-size scaling arguments \cite{FdG78} and renormalization-group analyses \cite{KD91,KD92a,Die86a}, it should take the scaling form
\begin{equation}\label{eq:FCscalf}
\beta \mathcal{F}_\mathrm{C}(T,L) \aseq L^{-d}\,\vartheta(x),\quad x\equiv t(L/\xi_+)^{1/\nu}
\end{equation}
in the scaling limit $t\equiv T/T_{\mathrm{c}}-1\to 0$ and $L\to \infty$, with $x$ kept fixed. Here we used $\xi_+$, the nonuniversal amplitude of the bulk correlation length $\xi(t>0)\aseq \xi_+t^{-\nu}$ in the disordered phase, to fix the scale of $x$. The scaling function $\vartheta(x)$ is universal; it depends only on gross properties of the medium, boundaries, and geometry (bulk and surface universality classes, large-scale boundary conditions, etc), but not on microscopic details. For the $O(2)$ case of a slab confined by free surfaces, information about $\vartheta(x)$ is available from two sources: from experiments on the thinning of $_{}^4$He wetting films on copper substrates as $T$ is lowered below the $\lambda$~transition  \cite{GC99,GSGC06}, and from Monte Carlo calculations for $XY$ models  on  simple cubic lattices with uniform nearest-neighbor interactions \cite{Huc07,VGMD07,Hasenbusch0905}. Both consistently indicate that $\vartheta(x)$ has the following properties:

\begin{enumerate}

\item[(p1)] it is negative, and hence $\beta \mathcal{F}_\mathrm{C}(T,L)$ attractive, for all $x$; this property also follows from the rigorous theorem for reflection positive systems in a slab geometry with symmetric boundary conditions proved in  \cite{Bac07} (for an analogous theorem for Gaussian models, see \cite{KK06}).

\item[(p2)] it decreases $\propto \exp(-\mathit{const}\,x)$ for $x\gg 1$; this property is in conformity with renormalization-group improved perturbation theory \cite{KD91} and ultimately follows from the exponential decay of correlation functions in the disordered phase.

\item[(p3)] its $t=0$ critical value $\vartheta(0)$ is rather small.

\item[(p4)]  it has a deep smooth minimum $\vartheta_{\mathrm{min}}$ located at $x_{\mathrm{min}}<0$.

\item[(p5)] it approaches a zero-temperature limit $\vartheta(-\infty)<0$.

\end{enumerate}

Standard perturbative renormalization-group approaches based on the $\epsilon=4-d$ expansion reproduce only some of these properties, yet fail to yield others such as (p4) and (p5). Their first application to the study of thermodynamic Casimir forces was restricted to the critical point and Dirichlet boundary conditions at both surface  planes (D-D boundary conditions) \cite{Sym81}. Then two-loop calculations giving the $\epsilon$~expansions of the scaling functions $\vartheta(x)$ to $O(\epsilon)$ in the paramagnetic phase $t>0$ for five different boundary conditions followed \cite{KD91,KD92a}. In addition to periodic (pbc), antiperiodic, and D-D boundary conditions, also special  (sp) boundary conditions of the Robin type corresponding to the critical enhancement of the surface interactions either on both surface planes (sp-sp) or on just one of them were considered, namely, the combinations sp-D and D-sp with Dirichlet boundary conditions on the respective second plane. The results of \cite{KD91,KD92a} for periodic and sp-sp boundary conditions hinted at problems with the $\epsilon$~expansion at $t=0$. Subsequent work \cite{DGS06,GD08,DG09} revealed that the  $\epsilon$~expansions of the Casimir forces at $t=0$ actually break down for these boundary conditions. The origin of the problem may be traced back to the presence of a zero mode at the bulk critical temperature in the Ornstein-Zernike (zero-loop) approximation, which thus predicts a sharp transition for finite $L$ there. The associated infrared singularities imply that the conventional RG-improved perturbation theory becomes ill-defined at $t=0$. Appropriate modifications of it have shown that the small-$\epsilon$ expansions of both $\vartheta^{\text{pbc}}(0)$ and $\vartheta^{\text{sp-sp}}(0)$  involve fractional powers $\epsilon^{k/2}$ with $2\le k\in \mathbb{N}$, modulated by powers of $\ln\epsilon$ when $k\ge 5$ \cite{DGS06,pbc,Sac97,GD08,SD08,DS11}. This breakdown of the $\epsilon$~expansion at $t=0$ for periodic and sp-sp boundary conditions makes extrapolations based on  small-$\epsilon$ expansions to low orders rather unreliable. 

In the case of D-D boundary conditions the situation is somewhat better. The Ornstein-Zernike approximation yields a zero-mode at a shifted temperature $t^{\text{D-D}}(L)<0$, and hence predicts a sharp transition for finite $L$ there. Thus the $\epsilon$~expansion remains valid for all  $t\ge 0$. This applies more generally for Robin boundary conditions corresponding to subcritical enhancement of the surface interactions \cite{DS11}, where Ornstein-Zernike theory yields a zero-mode and hence  a sharp finite-$L$ transition at a shifted temperature in the open interval $(t^{\text{D-D}}(L),0)$. These predictions of sharp $L<\infty$  transitions below the bulk critical temperature are qualitatively correct for the $({d=3})$-dimensional scalar ($n=1$) case. The main hard challenge one is faced with is to design a theory that is capable of handling in addition to the infrared singularities at $t=0$ also those at the shifted critical temperature and the associated dimensional crossover. 

The situation is worse in the $({d=3})$-dimensional $O({n\ge 2})$ case. Since long-range order is rigorously ruled out for finite thickness $L$ at all temperatures $T>0$ by the Mermin-Wagner theorem \cite{MW66,MW94}, only a rounded $L<\infty$ transition is possible when $T>0$, where the $O(2)$ case is special in that a Kosterlitz-Thouless transition to a low-temperature phase with quasi-long-range order is known to occur at a nonzero temperature $T_{\text{KT}}(L)<T_{\mathrm{c}}$ (see \cite{Hasenbusch0902} and its references). The destruction of long-range order at low temperatures caused by low-energy fluctuations is a nonperturbative phenomenon (``nonperturbative mass generation''). Any  theory aiming at a satisfactory description of the Casimir force for the full range $(-\infty,\infty)$ of the scaling variable $x$ must be able to account for it. Otherwise the scaling function  $\vartheta(x)$ it yields cannot even qualitatively be correct.

Given these challenges and the present unsatisfactory state of the theory, reliable knowledge from exact solutions
of appropriate model systems could be extremely useful: It could serve a dual purpose,  providing both a starting point and benchmark for approximate treatments of more realistic models. Exact solutions of $O(n)$ models in the limit $n\to\infty$ lend themselves to these goals because fluctuations can be dealt with in a systematic, mathematically controlled fashion when $n$ becomes large. This applies to both critical and Goldstone mode fluctuations. Furthermore, the theory succeeds in generating a nonzero mass for $T>0$ in two-dimensional bulk systems \cite{largen,MZ03}. 

The usefulness of exact $n\to\infty$ results for fluctuation-induced forces of $O(n)$ models on (${d=3}$)-dimensional films  has been convincingly demonstrated for the case of periodic boundary conditions. Danchev\cite{Dan96,Dan98} managed to compute the thermodynamic Casimir force as a function of $T$ and magnetic field  $h$ in the limit $n\to\infty$. The behavior of $\vartheta^{\mathrm{pbc}}(x)$, the  analog  of the scaling function $\vartheta(x)$ introduced in Eq.~\eqref{eq:FCscalf}, in the vicinity of $T_{\mathrm{c}}$ gave clear indications of problems with the $\epsilon$~expansion for pbc in this temperature regime. The exact $n=\infty$ critical value $\vartheta^{\mathrm{pbc}}(0)=-4\zeta(3)/5\pi\simeq -0.306$ turned out to be fairly close to the Monte Carlo value $\simeq-0.304$  for the Ising case $n=1$ \cite{Kre97,DK04,VGMD09,HGS11}. However, evaluating Krech and Dietrich's $O(\epsilon)$ results  \cite{KD91,KD92a} at $\epsilon=1$ for $n=1,2,3,\infty$ gives values of $\vartheta^{\mathrm{pbc}}(0)$ that deviate strongly from the exact $n=\infty$~result for $d=3$. Even worse, the differences increase as $n$ grows \cite{BDT00,GD08}.

Subsequent work revealed that the $\epsilon$~expansion actually breaks down at $T_{\mathrm{c}}$ for pbc \cite{Sac97,DGS06,GD08}. Thus the exact $n=\infty$ results for the $({d=3})$-dimensional case with pbc have provided helpful guidance and a benchmark for assessing  the quality of estimates based on both the $\epsilon$~expansion and Monte Carlo calculations for finite values of $n$. A similarly important role have exact $n\to\infty$ solutions played in the theory of crossover behavior near quantum critical points \cite{Sac11}.  In fact, close analogies exist between classical models on a strip with pbc and bosonic quantum systems  in $2+1$ spacetime dimensions whose dynamic critical exponent is $z=1$. This follows from the well-known fact that the latter at temperature $T=1/k_{\mathrm{B}}\beta$ can be mapped onto (${d=3}$)-dimensional classical systems on a strip of width $\beta$ subject to  pbc along the $\beta$~direction.

On the other hand, the exact $n=\infty$  scaling function $\vartheta^{\mathrm{pbc}}(x)$ for $d=3$ does not exhibit a local minimum below $T_{\mathrm{c}}$ (property (p4)). It rather decreases smoothly and monotonically from its maximum value zero at  temperatures above $T_{\mathrm{c}}$ to its Goldstone value $\vartheta^{\mathrm{pbc}}(-\infty)=-\zeta(3)/\pi$ \cite{Dan96,Dan98,Sac93}, where it saturates. In order for the Casimir force to have a local minimum at $T<T_{\mathrm{c}}$, free boundary conditions and the implied breaking of translational invariance along the $z$~direction appear to be crucial.

The purpose of the present paper is to compute the scaling function $\vartheta(x)$ and its counterpart for the excess free energy for free boundary conditions and $d=3$ exactly in the limit $n\to\infty$. Owing to these boundary conditions, translation invariance is broken along the $z$~direction. This implies that the $n\to\infty$ limit is not given by the solution of a mean spherical model \cite{Sta71,BDT00} with a global constraint on the sum $\sum_i\langle s_i^2\rangle$ of the expectation values of the square of spin variables over all sites $i$. Instead, separate constraints of this kind must be imposed on the respective sums $\sum_{i\in z}\langle s_i^2\rangle$ for each layer $z$ \cite{Kno73,BM77a,BM77c}. The associated $z$-dependent Lagrange multipliers correspond to a quadratic interaction $V(z)\,\phi^2$, where the potential $V(z)$ must be determined self-consistently by solving the constraint equations along with a Schrödinger equation (see, e.g., \cite{BM77a,BM77c,DADC08,CHG09}, \cite[Appendix B]{BDS10}, and \cite{DGHHRS12}). Bray and Moore \cite{BM77a,BM77c} succeeded in determining the solution $V(z)$ in the scaling regime  for the special case  of a semi-infinite system at bulk criticality, $L=\infty$, $t=0$, in closed analytic form. Whether the self-consistent potential $V(z)$ or even the spectrum and eigenfunctions of  the corresponding Schrödinger equation can also be obtained in analytical closed form for finite $L$ and away from $T_{\mathrm{c}}$ is not at all clear, if not unlikely. We therefore attack these problems below by numerical means. 

The remainder of the paper is organized as follows. In the next section, we introduce the continuum $\phi^4$ model on a slab whose large-scale behavior we are going to study. We begin with general considerations concerning the corrections to scaling  that can be expected on general grounds for the critical Casimir force in $d=3$ dimensions. Since the correction-to-scaling exponent $\omega$ of the Wegner bulk corrections takes the exact $n=\infty$ value $\epsilon\equiv 4-d$, it becomes $\omega=1$ at $d=3$. However, in systems bounded by ($d{-}1$)-dimensional surface planes one expects quite generally irrelevant surface scaling fields that scale as a length \cite{DDE83,Die86a,Die97}. Since these two types of irrelevant scaling fields become degenerate at $d=3$, logarithmic anomalies occur in surface and finite-size quantities such as the Casimir force, as will be explained in Sec.~\ref{sec:corrtoscal}. 

For our subsequent numerical analysis of  the self-consistent Schrödinger equation that the exact  $n\to\infty$ solution involves, a discretization of our model is necessary. We use two distinct discretization schemes: In the first (Sec.~\ref{sec:pdmod}), we merely discretize along the $z$~direction, leaving continuous the coordinates along the other (``parallel'') directions, and using dimensional regularization to regulate the ultraviolet singularities (UV) of the required parallel momentum integrations.  We then show how the Schrödinger equation involving the discretized version of the operator $-\partial_z^2$ can be efficiently solved at $T_{\mathrm{c}}$. The convergence of the solution depends significantly on the value of the $\phi^4$ interaction constant $g$. By taking an appropriate $g\to\infty$ limit, we manage to obtain simplified equations, improve the speed of convergence, and suppress logarithmic corrections. Subsequently, the analysis is extended to temperatures $T\ne T_{\mathrm{c}}$. Precise results for the Casimir amplitude  and the scaling functions  of the excess free energy and the Casimir force are derived for the case of asymptotic Dirichlet boundary conditions. 

The use of partial discretization in conjunction with dimensional regularization means that not all corrections to scaling due to a finite lattice constant are incorporated. This prompted us to check and corroborate our findings by a separate careful study of a fully discretized model. The corresponding lattice model and its analysis is described in Sec.~\ref{sec:lattmod}. Taking again an appropriate $g\to\infty$ ($T_{\mathrm{c}}\to\infty$) limit, we are able to make contact with the simplified equations of Sec.~\ref{sec:pdmod} in which corrections to scaling are suppressed. 

Logarithmic anomalies  manifest themselves also in the low-temperature behavior. They are produced by  Goldstone-mode excitations on length scales smaller than the Josephson correlation length. To gain information about their effects on the behavior of the Casimir force scaling function $\vartheta(x)$ in the limit $x\to-\infty$, we use the fact that our $O(n)$ $\phi^4$ film model with free boundary conditions can be mapped at low temperatures onto a nonlinear sigma model. This mapping is expounded in Appendix~\ref{sec:derivnlsigma} and exploited in Appendix~\ref{sec:lowtas} to determine the asymptotic form of $\vartheta(x)$ as $x\to-\infty$, which turns out to involve logarithmic anomalies. Our main findings are stated at the end of Sec.~\ref{sec:pdmod}. A more detailed analytical investigation of the low-temperature asymptotics of the scaling functions is reserved for a subsequent paper \cite{RD13}.

In Sec.~\ref{sec:exprop}, we gather the available knowledge about exact properties that is relevant for our subsequent numerical work. Section~\ref{sec:numan} then follows with a detailed account of our methods used to determine the numerical solutions of the self-consistent equations for both the partially discretized and the lattice model and a presentation of their results. Our high-precision data for the lattice model turn out to agree to all significant digits with those for the partially discretized one. Section~\ref{sec:sumconc} contains a brief summary of our results and our conclusions. In addition to the two Appendices~\ref{sec:derivnlsigma} and \ref{sec:lowtas} already mentioned, there is a third one (Appendix~\ref{sec:fex})  to which some technical details have been relegated.

\section{Continuum model and large-component limit} \label{sec:contmodlnl}

\subsection{Continuum model}\label{sec:contmod}

A standard continuum model for studying critical behavior of a $d$-dimensional strip $\mathfrak{V}=\mathbb{R}^{d-1}\times [0,L]$ bounded by two free surfaces at $z=0$ and $z=L$  in the absence of symmetry-breaking fields is defined by the $O(n)$-symmetrical Hamiltonian
\begin{align}\label{eq:Ham}
\mathcal{H}=&\int_{\mathfrak{V}}\!\rmd^dx\;\Big[\frac{1}{2}(\nabla\bm{\phi})^2+\frac{\mathring\tau}{2}\phi^2+\frac{g}{4! n}\phi^4\Big]\nonumber\\ &+\sum_{j=1}^2\int_{\mathfrak{B}_j}\!\rmd^{d-1}y\;\frac{\mathring c_j}{2}\phi^2\;.
\end{align}
Here $\bm{\phi}=(\phi_a,a=1,\dotsc,n)$ is an $n$-component order-parameter field, and the  usual short hand $(\nabla\bm{\phi})^2=\sum_{a=1}^n(\nabla\phi_a)^2$ is used. We write the position vector as $\bm{x}=(\bm{y},z)$, decomposing it into a $(d{-}1)$-dimensional coordinate $\bm{y}$ parallel to the surface planes $\mathfrak{B}_1=\{(\bm{y},0)\mid\bm{y}\in\mathbb{R}^{d-1}\}$ and $\mathfrak{B}_2=\{(\bm{y},L)\mid\bm{y}\in\mathbb{R}^{d-1}\}$ and a one-dimensional coordinate $z$ perpendicular to them. Since we wish to study this model in the limit $n\to\infty$, we normalized the $\phi^4$ interaction constant such that the limit can be taken at fixed $g$. 

Let
\begin{equation}\label{eq:Z}
\mathcal{Z}=\int\mathcal{D}[\bm{\phi}]\,\rme^{-\mathcal{H}[\bm{\phi}]}
\end{equation}
be the partition function of this model. We wish to determine the reduced free energy per base area $A=\int_{\mathbb{R}^{d-1}}\rmd^{d-1}y$ of the slab and number of components,
\begin{equation}\label{eq:fLdef}
f_L=-\lim_{n\to\infty}\frac{\ln\mathcal{Z}}{nA}\;.
\end{equation}
For later use, let us also introduce the correspondingly defined reduced bulk free energy density
\begin{equation}\label{eq:fbdef}
\fb=\lim_{L\to\infty}f_L/L
\end{equation}
and the reduced excess free energy density
\begin{equation}\label{eq:fexdef}
\fex(L)\equiv f_L-L\,\fb,
\end{equation}
whose limiting value 
\begin{equation}\label{eq:fexfs}
\fex(\infty)\equiv\lim_{L\to\infty}\fex(L)=f_{\mathrm{s}}=f_{\mathrm{s},1}+f_{\mathrm{s},2}
\end{equation}
yields the sum $f_{\mathrm{s}}$ of the surface free energy densities $f_{\mathrm{s},1}$ and $f_{\mathrm{s},2}$ of the two semi-infinite systems with the boundary planes $\mathfrak{B}_1$ and $\mathfrak{B}_2$, respectively. We will refer to the difference
\begin{equation}\label{eq:fresdef}
f_{\mathrm{res}}(L)=\fex(L)-f_{\mathrm{s}}
\end{equation}
as residual free energy. 

In the scaling regime (small $|T/T_{\mathrm{c}}-1|$, large $L$), this quantity is expected to have a scaling form analogous to Eq.~\eqref{eq:FCscalf}, namely (see, e.g., \cite{KD91})
\begin{equation}\label{eq:fresscf}
f_{\mathrm{res}}(T,L)\aseq L^{-(d-1)}\,\Theta(x)\,,
\end{equation}
from which the Casimir force
\begin{equation}\label{eq:betaFCdef}
\beta \mathcal{F}_{\mathrm{C}}=-\frac{\partial}{\partial L}f_{\mathrm{res}}(L)=-\frac{\partial}{\partial L}\fex(L)
\end{equation}
can be computed in a straightforward fashion to conclude that
\begin{equation}\label{eq:relvarthetaTheta}
\vartheta(x)=(d-1)\,\Theta(x)-\frac{x}{\nu}\,\Theta'(x),
\end{equation}
while the Casimir amplitude $\D{}$ is defined as 
\begin{equation} \label{eq:DeltaCdef}
\D{} \equiv \Theta(0).
\end{equation}

The relevant large-$n$ equation from which the above quantities  are to be computed can be  derived by standard methods (see, e.g., \cite{Eme75,MZ03} and \cite[Appendix B]{BDS10}). Introducing an auxiliary field $\psi(\bm{x})$, we can make a Hubbard-Stratonovich transformation 
\begin{equation}\label{eq:HS}
\rme^{-\frac{g}{4! n}\phi^4}=\sqrt{\frac{3n}{2\pi g}}\int_{-\infty}^{\infty}\rmd\psi\,\rme^{\frac{1}{2}\phi^2\,\rmi\psi-\frac{3n}{2g}\,\psi^2}
\end{equation}
to obtain
\begin{equation}\label{eq:Z_HS}
\mathcal{Z}=C\int\mathcal{D}[\bm{\phi}]\int\mathcal{D}[\psi]\,\rme^{-\frac{1}{2}\int\rmd^{d}x\big[\bm{\phi}\left(-\nabla^{2}+\tb+
\rmi\psi\right)\bm{\phi}+\frac{3n}{g}\psi^{2}\big]}\,,
\end{equation}
where $C$ is a constant (depending on $g/n$). To arrive at the derivative term of the action  in Eq.~\eqref{eq:Z_HS} we integrated by parts. The boundary terms produced by this operation cancel those resulting  from the surface integrals $\int_{\mathfrak{B}_j}$ of the Hamiltonian \eqref{eq:Ham} provided the Robin boundary conditions (cf., for example, \cite{Die86a,SD08,DS11}),
\begin{align}\label{eq:bc}
(\partial_z-\cb_1)\bm{\phi}(\bm{y},0)&=0,\nonumber\\
(\partial_z+\cb_2)\bm{\phi}(\bm{y},L)&=0,
\end{align}
hold. The Laplacian must be interpreted accordingly; with these boundary conditions imposed, it is self-adjoint.

\subsection{Large-$n$ limit}\label{sec:lnl}
The large-$n$ behavior of the functional integral~\eqref{eq:Z_HS} follows via a saddle-point integration. Since translation invariance is broken along the $z$~direction, we must look for a $z$-dependent extremum $\psi\equiv\psi(z)$. It is convenient to express this as 
\begin{equation}\label{eq:pot}
\rmi\psi(z)=V(z)-\tb
\end{equation}
in terms of a potential $V(z)$. Let us restrict ourselves to the case of disordered  phases (with unbroken $O(n)$ symmetry). Then we can integrate out the  order-parameter field $\bm{\phi}$ in a straightforward fashion. Upon taking a Fourier transform with respect to the $\bm{y}$ coordinate, we arrive at
\begin{align}\label{eq:f}
f_L={}&\frac{1}{2}\pint\Tr{\big[}\ln(\bm{p}^{2}-\partial_{z}^{2}+V)\big]\nonumber \\
 &-\frac{3}{2g}\int_{0}^{L}\rmd{z}\,{[\tb-V(z)]}^{2}+f_L^{(0)}\,,
\end{align}
where $f_L^{(0)}$ is a trivial background term which we shall drop henceforth since it does not affect the universal quantities we are concerned with. Here, the Dirac notation  $\Tr(\dots) = \int_{0}^{L}\mathrm{d}z\langle z|\dots |z\rangle$ and  the short hand
\begin{equation}\label{eq:pint}
\pint\equiv \int_{-\infty}^{\infty}\frac{\rmd^{d-1}p}{(2\pi)^{d-1}}
\end{equation}
are used. Just as the Laplacian, the operator $-\partial_z^2$ is subject to the boundary conditions~\eqref{eq:bc}.

The stationarity of $f_L$ at $V(z)$ implies the condition
\begin{align}\label{eq:dfdV}
\frac{\delta f_L}{\delta V(z)}={}&\frac{1}{2}\pint\langle z|{[\bm{p}^{2}-\partial_{z}^{2}+V]}^{-1}|z\rangle\nonumber\\
 &{} +\frac{3}{g}[\tb-V(z)]=0.
\end{align}
This is a nontrivial equation for $V(z)$, which can be cast in a more convenient form by introducing 
 a complete orthonormal set of eigenfunctions $\{\varphi_\nu(z)=\langle z|\nu\rangle\}$ satisfying
\begin{equation}\label{eq:ev}
[-\partial_{z}^{2}+V(z)]\varphi_{\nu}(z)=\varepsilon_{\nu}\varphi_{\nu}(z)
\end{equation}
along with the boundary conditions~\eqref{eq:bc}. Using these eigenfunctions, we can solve Eq.~\eqref{eq:dfdV} for $\tb-V(z)$ to obtain
\begin{equation}\label{eq:V-tau}
\tb-V(z)=-\frac{g}{6}\pint\sum_{\nu}\frac{|\varphi_{\nu}(z)|^2}{\bm{p}^{2}+\varepsilon_{\nu}}.
\end{equation}
Equation~\eqref{eq:V-tau} for the potential  and the  Euclidean Schrödinger equation~\eqref{eq:ev}, together with the boundary conditions~\eqref{eq:bc}, form a set of equations that must be solved self-consistently for $V(z)$ and the eigenfunctions $\varphi_\nu(z)$.

\subsection{Remarks}\label{sec:rem}

Nonclassical bulk critical behavior is known to occur for dimensions $d$ between the upper and lower bulk critical dimensions, i.e., for $2<d<4$. Our primary concern in this paper is to determine solutions to the above equations for $d=3$. Let us nevertheless temporarily consider the more general case $2<d<4$. Several remarks about the above equations \eqref{eq:f}, \eqref{eq:ev}, and \eqref{eq:V-tau} are necessary. 

The first concerns the UV behavior of the required momentum integrals. The integrals $\pint$ of individual summands labeled by $\nu$ in Eq.~\eqref{eq:V-tau} are not guaranteed to be UV convergent when $d\ge 3$. If we regularize them by restricting the integration to $|\bm{p}|\le \Lambda$, power counting tells us that they vary as $\Lambda^{d-3}$. We must also take into account that the mode summation $\sum_\nu$ in the limit $L\to\infty$ involves an integration over a set of one-dimensional wave vectors $0\le k<\infty$. Hence a leading UV singularity $\sim\Lambda^{d-2}$ is to be expected. We can get rid of  the UV divergence in Eq.~\eqref{eq:V-tau} by subtracting from this equation its bulk analog at the bulk critical point $T_{\mathrm{c}}$. To understand this, it
 will be helpful to see how  information about the bulk case can be recovered from the above self-consistent equations. Taking the limit $L\to\infty$ gives us a semi-finite system. Let us denote the potential $V(z)\equiv V(z|L)$  for this case as $V_\infty(z)\equiv V(z|\infty)$. As $z\to\infty$, this potential must approach the bulk value, which is nothing but the inverse $r_{\mathrm{b}}$ of the bulk susceptibility $\chi_{\mathrm{b}}$:
\begin{equation}\label{eq:Vbulk}
\lim_{z\to\infty}V_\infty(z)= V_\infty(\infty)=r_{\mathrm{b}}.
\end{equation}
The bulk analogs of the eigenvalues $\varepsilon_\nu$ are  continuous functions $\varepsilon_{\mathrm{b}}(k)$ of the wavenumber $k$ conjugate to $z$. From Eq.~\eqref{eq:Vbulk} and the large-$z$ limit of Eq.~\eqref{eq:ev} we see that they are given by
\begin{equation}\label{eq:epsbk}
\varepsilon_{\mathrm{b}}(k)=r_{\mathrm{b}}+k^2,\quad k\in(0,\infty),
\end{equation}
for our continuum model~\eqref{eq:Ham}.

It follows from these results in conjunction with  Eq.~\eqref{eq:V-tau} that the bulk critical point is located at
\begin{equation}\label{eq:tbc}
\tb_\mathrm{c}=-\frac{g}{6}\pint\int_{k>0}\frac{\rmd k}{\pi}\frac{1}{\bm{p}^2+\varepsilon_{\mathrm{b},\mathrm{c}}(k)}\;,
\end{equation}
where $\varepsilon_{\mathrm{b},\mathrm{c}}(k)$ means the critical ($r_{\mathrm{b}}=0$) analog of $\varepsilon_{\mathrm{b}}(k)$. Our reason for writing $\varepsilon_{\mathrm{b},\mathrm{c}}(k)$ rather than $k^2$ is to prepare for our analysis below that uses a discretization along the $z$~direction in conjunction with dimensional regularization of the $\bm{p}$~integrations. For simplicity, we take both the nearest-neighbor (NN) bond and the lattice constant along the $z$~direction to be unity.  Then Eq.~\eqref{eq:tbc} remains valid in the given form except that the corresponding linear-chain dispersion relation 
\begin{equation}\label{eq:epsmclat}
\varepsilon_{\mathrm{b,c}}^{\mathrm{lc}}(k)=4\sin^2(k/2),\quad 0\le k\le \pi,
\end{equation}
must be substituted for $\varepsilon_{\mathrm{b},\mathrm{c}}(k)$ and the $k$ integration restricted to the interval $(0,\pi)$.

Upon setting
\begin{equation}
\tb=\tb_\mathrm{c}+\tau,
\end{equation}
we can now subtract  Eq.~\eqref{eq:tbc} from Eq.~\eqref{eq:V-tau} to obtain
\begin{subequations}\label{eq:V-dt-d-p}
\begin{equation}\label{eq:eqs}
\tau-V(z)=-\frac{g}{6}\,I_L(z)
\end{equation}
with 
\begin{equation}\label{eq:ILz}
I_L(z)=
\pint\bigg[\sum_{\nu}\frac{|\varphi_{\nu}(z)|^2}{\bm{p}^{2}+\varepsilon_{\nu}}-\int_{k>0}\frac{\rmd k}{\pi}\frac{1}{\bm{p}^{2}+\varepsilon_{\mathrm{b},\mathrm{c}}(k)}\bigg]. 
\end{equation}
\end{subequations}
The subtraction provided by the second term in square brackets removes the leading UV singularity. This is evident for the bulk case of our continuum model where $I_L(z)$ becomes 
\begin{equation}\label{eq:Ib}
I_\infty(\infty)\equiv I_{\mathrm{b}}= -\int_{\bm{q}}^{(d)}\frac{r_{\mathrm{b}}}{q^2(q^2+r_{\mathrm{b}})}
\end{equation}
and UV convergent for $d<4$.

The UV finiteness of $I_\infty(z)$ can be explicitly verified both for the semi-infinite case  \cite{BM77a,BM77c} and  that of pbc \cite{DDG06}. We refrain from an explicit demonstration of the UV finiteness of $I_L(z)$ when $L<\infty$ for the fully continuous model~\eqref{eq:Ham} since some sort of discretization will be needed for the numerical analysis of the above self-consistency equations. In Sec.~\ref{sec:pdmod} we shall explicitly show that a discretization along the $z$~direction is sufficient to render the analog of the difference on the right-hand side of Eq.~\eqref{eq:ILz} UV finite.  Thus no UV cutoff is needed to deal with the set of self-consistent equations~\eqref{eq:ev} and \eqref{eq:V-dt-d-p}. However, the UV behavior of contributions to $f_L$ is 
worse. Therefore, appropriate subtractions are necessary to obtain UV finite differences (see Sec.~\ref{sec:pdmod}). In the case of the lattice discretization used in Sec.~\ref{sec:lattmod}, the UV convergence of quantities such as $I_L(z)$ and bulk, surface, and excess free energies is, of course,  trivially ensured because  the wave vectors $\bm{q}$ are restricted to the first Brillouin zone.

Our second remark concerns the challenge of finding exact solutions to Eqs.~\eqref{eq:ev} and \eqref{eq:V-dt-d-p}. This is straightforward in the bulk case because of translation invariance, but nontrivial already for semi-infinite systems. Bray and Moore~\cite{BM77a,BM77c} succeeded in determining the exact large-scale forms of the potentials $V_{\infty,\mathrm{c}}(z)\equiv V(z|t{=}0,L{=}\infty)$ at the bulk critical point. For $3<d<4$, they found two solutions, namely
\begin{subequations}\label{eq:BMV}
\begin{equation}\label{eq:BMVord}
V^{\text{ord}}_{\infty,\mathrm{c}}(z)
=\frac{(d-3)^2-1}{4z^2}\quad(\text{for }2<d<4)
\end{equation}
and 
\begin{equation}\label{eq:BMVsp}
V^{\text{sp}}_{\infty,\mathrm{c}}(z)
=\frac{(5-d)^2-1}{4z^2}\quad(\text{for }3<d<4),
\end{equation}
\end{subequations}
associated with the ordinary and special surface transitions, respectively. For $2<d\le 3$, only $V^{\text{ord}}_{\infty,\mathrm{c}}(z)$ remains. No exact solutions $V_{\infty}(z)$ away from bulk criticality are known in closed analytical form. Whether Bray and Moore's results can be generalized so as to determine the exact self-consistent potential $V_{L,\mathrm{c}}(z)$ in closed analytical form for finite $L$, either just at $t=0$ or even at $t\ne 0$, is unclear to us and appears  to be an extremely difficult problem to which we have at present no solution. We will therefore resort to numerical methods below.

Our third remark concerns the phase behavior of the $n=\infty$ model. For finite $L$, it should behave as an effective $(d{-}1)$-dimensional system on sufficiently long length scales. As pointed out already in the introduction, the Mermin-Wagner theorem \cite{MW66} precludes a phase with long-range order at $d=3$ when $L<\infty$. Likewise, no long-range ordered surface phase can occur in the semi-infinite case when $d\le3$. This means that in our analysis of the (${d=3}$)-dimensional case, only the solution~\eqref{eq:BMVord} pertaining to the ordinary transition must be considered.

\section{Corrections to scaling}\label{sec:corrtoscal}

For precise numerical determinations of scaling functions detailed knowledge of corrections to scaling is essential.  Anomalous corrections to scaling must be expected for surface and finite-size quantities on general grounds at $d=3$ when $n=\infty$.

We begin by recalling the dependence of the bulk integral~\eqref{eq:Ib} on $r_{\mathrm{b}}$ when the momentum integration is cut off by means of a $\Lambda$-dependent cutoff function. This is analyzed for a general class of cutoff functions in \cite[Appendix A]{MZ03}. The result is that $I_\mathrm b$ behaves as
\begin{align}\label{eq:Ibas}
I_{\mathrm{b}}(r_{\mathrm{b}})&=\Lambda^{d-2}\,I_{\mathrm{b}}\big(x^2=r_{\mathrm{b}}/\Lambda^2\big)\nonumber\\
&=\Lambda^{d-2}\left[-A_d\,x^{d-2}+w_d\,x^2+O(x^4,x^d)\right]
\end{align}
where
\begin{equation}\label{eq:Ad}
A_d=-(4\pi)^{-d/2}\,\Gamma(1-d/2)
\end{equation}
is a universal coefficient (independent of the chosen regularization). By contrast, $w_d$ is nonuniversal (regularization dependent). It can have either sign for given $d\in(2,4)$, yet has a pole term at $d=4$ with the same residue as $A_d$:
\begin{equation}
w_{4-\epsilon}\mathop{=}_{\epsilon\to 0}\frac{1}{8\pi^2\epsilon}+O(\epsilon^0)=A_{4-\epsilon}+O(\epsilon^0).
\end{equation}

As examples, we give the values of $w_3$ for the following three distinct kinds of regularizations:

\begin{enumerate}
\item[(a)] a sharp cutoff regularization; this means that the integration $\int_{\bm{q}}^{(d)}$ in Eq.~\eqref{eq:Ib} is restricted to the $d$-ball $|\bm{q}|\le\Lambda$.

\item[(b)] the mentioned discretization of  the $z$~coordinate, combined with  dimensional regularization of the parallel momentum integrations $\pint$. 

\item[(c)] introduction of a simple cubic lattice (lattice constant $a=1$, NN bond $=1$). 

\end{enumerate}

In case (c), the analog of Eq.~\eqref{eq:Ib} can be expressed as a difference
\begin{equation}\label{eq:Iblat}
I_{\mathrm{b}}(r_{\mathrm{b}})=W_3(r_{\mathrm{b}})-W_3(0)
\end{equation}
of standard Watson integrals  defined by \cite{JZ01}
\begin{equation}\label{eq:Wddef}
W_d(\lambda)\equiv\int_0^\pi\frac{\rmd{q}_1}{\pi}\dotsm\int_0^\pi\frac{\rmd{q}_d}{\pi}\frac{1}{\lambda+4\sum_{i=1}^d\sin^2(\frac{q_i}{2})}.
\end{equation}

These regularizations (a)--(c) yield the values
\begin{equation}\label{eq:w3}
w_3=\begin{cases}
w_3^{(a)}=(2\pi^2)^{-1},\\
w_3^{(b)}=0,\\
w_3^{(c)}=-0.012164158583\dots\,.
\end{cases}
\end{equation}
The calculation of $w_3^{(a)}$ is elementary, $w_3^{(b)}$ follows by dimensional arguments, and $w_3^{(c)}$ may be obtained in exact analytical form from the results for Watson integrals given in \cite{Joy01}. They yield 
\begin{equation} \label{eq:w3c}
w_3^{(c)} = \frac{1}{64\pi^2 W_3(0)} - \frac{7 W_3(0)}{96}
\end{equation}
with \cite{JZ01}
\begin{align} \label{eq:W30}
W_3(0)&=\frac{\sqrt{3}-1}{192 \pi^3}
\left[\Gamma\left(\frac{1}{24}\right)
  \Gamma\left(\frac{11}{24} \right)
\right]^2 \nonumber \\
&=0.252731009858663\dots\,.
\end{align}

These results exemplify the known fact that the coefficient $w_d$ can have either sign or vanish. Since an adequate discussion of the role of $w_d$ and the issue of its sign can be found in \cite[p.~87]{MZ03}, we can be brief. The important point is that whenever $w_d>0$, the value $g^*=6\Lambda^\epsilon/w_d$ of $g$ may be interpreted as the location of an infrared-stable fixed point if $2<d<4$. It will be sufficient for our purposes to verify the consistency of this statement with the corrections to scaling the large-$n$ solution yields for the bulk susceptibility. By combining  Eqs.~\eqref{eq:Vbulk}, \eqref{eq:eqs}, \eqref{eq:Ib}, and \eqref{eq:Ibas}, we recover a familiar result for the bulk equation of state,
namely
\begin{equation}
\tau/ r_{\mathrm{b}}=1-\frac{g\Lambda^{-\epsilon}}{6/w_d}+\frac{g}{6}\,A_d\,r_{\mathrm{b}}^{-\epsilon/2}+O\big(r_{\mathrm{b}}^{1-\epsilon/2}/\Lambda^2\big).
\end{equation}
Its solution for small $\tau$ and $r_{\mathrm{b}}$ gives
\begin{equation}\label{eq:tauchi}
\tau \aseq \frac{gA_d}{6}\,r_{\mathrm{b}}^{1-\epsilon/2}\left[1+\frac{6}{g A_d}\bigg(1-\frac{g\Lambda^{-\epsilon}}{6/w_d}\bigg)r_{\mathrm{b}}^{\epsilon/2}\right]
\end{equation}
and enables us to read off the standard $n=\infty$ results
\begin{equation}\label{eq:gammaom}
\gamma=2\nu=\frac{2}{d-2},\qquad \omega=4-d,
\end{equation}
 for the bulk critical indices $\gamma$ and $\nu$ and the correction-to-scaling exponent $\omega$, respectively. 
 
If $w_d>0$ so that $g^*>0$, the corrections to scaling $\sim r_\mathrm{b}^{\epsilon/2}$ in Eq.~\eqref{eq:tauchi} can be eliminated by setting $g=g^*$, a trick used also in Bray and Moore's large-$n$ analysis  of the semi-infinite system at $t=0$ \cite {BM77a,BM77c}. However, when $w_d<0$, this is not possible since $g$ must be positive. To understand the limiting case $w_d=0$, it is helpful to consider a sequence of regularizations yielding positive values $w_d^{(j)}$, $j=1,2,\dots,\infty$, with  $\lim_{j\to\infty}w_d^{(j)}=0$. For any finite $j$, there is an infrared-stable fixed point whose location $g^*_j$ moves to $g^*=\infty$ as $j\to\infty$. As can be seen from Eq.~\eqref{eq:tauchi}, the corrections to scaling can still be suppressed by setting $g=g^*$ ($=\infty$) provided an appropriately scaled  temperature variable  $t=\mathit{const}\, \tau/g$ is introduced. This is the strategy we will employ in our analysis in Sec.~\ref{sec:pdmod}.

We next turn to the issue of corrections to scaling in the semi-infinite and film cases $L=\infty$ and $L<\infty$, respectively. Since we will mainly be concerned with the (${d=3}$)-dimensional situation, we can restrict ourselves to the ordinary transitions for which Dirichlet boundary conditions hold asymptotically on large scales \cite{Die86a,Die97}. It is well known that irrelevant surface scaling fields $\lambda_1$ and $\lambda_2$ associated with the boundaries $\mathfrak{B}_1$ and $\mathfrak{B}_2$ of semi-infinite systems exist which scale exactly as a length (a proof is given in \cite[Appendix C]{DDE83}). Physically, they correspond to so-called extrapolation lengths  which indicate the distance from the boundaries where the linear extrapolation of $\phi$ vanishes \cite{Die86a,Bin83,LR75}. An alternative way of understanding their presence is to note that the component $T_{zz}$ of the stress-energy tensor appears in the boundary operator expansion of the order parameter about $\mathfrak{B}_j$ \cite{Die97,EKD93}. Under RG transformations with a change $\mu\to\mu\ell$ of the momentum scale this operator scales as $\ell^{\Delta[T_{zz}]}$  with its engineering dimension $\Delta[T_{zz}]=d$. Since the RG-eigenexponent $y_\lambda$ of the scaling fields $\lambda_j$ and $\Delta[T_{zz}]$ must add up to the surface dimension $d-1$, we have $y_\lambda=-1$ and hence $\omega_\lambda=-y_\lambda=1$ for the associated correction-to-scaling exponent $\omega_\lambda$.

The result means that the $n=\infty$ correction-to-scaling exponents $\omega$ and $\omega_\lambda$ become degenerate at $d=3$. Such degeneracies are known to imply logarithmic anomalies. To show this, we can generalize Wegner's reasoning in \cite[Section V.E]{Weg76} in an appropriate fashion. Ignoring the above mentioned sign problem of $w_d$, we assume that a regularization has been chosen such that an infrared-stable fixed point with $g^*>0$ exists. Let $\delta g=\mu^{-\epsilon}(g-g^*)$ be the dimensionless linear scaling field \cite{remmu} associated with deviations of the dimensionfull coupling constant $g$ from its fixed-point value $g^*$,  and let $\check{\lambda}_j$ be the dimensionless linear surface scaling fields $\check{\lambda}_j=\lambda_j/\mu$. Just as any other bulk scaling field, $\delta g$ can be coupled  to other linear \emph{bulk} scaling fields in the flow equations, but not to any linear \emph{surface} scaling fields. By contrast, the surface scaling field $\check{\lambda}_j$ can be coupled to other linear bulk scaling fields as well as to surface scaling fields associated with the same surface plane $\mathfrak{B}_j$. Dropping all nonlinearities and ignoring couplings to other scaling fields, we arrive at phenomenological flow equations of the form
\begin{equation}
\ell\frac{\rmd}{\rmd\ell}\delta g(\ell)=\omega\,\delta g(\ell)+\dots
\end{equation}
and
\begin{equation}
\ell\frac{\rmd}{\rmd\ell}\check{\lambda}_j(\ell)=\check{\lambda}_j(\ell)+a_{j,g}\,\delta g(\ell)+\dots
\end{equation}
with the initial conditions $\delta g(1)=\delta g$ and $\check{\lambda}_j(1)=\mu\lambda_j$. Solving these equations gives the limiting large length-scale ($\ell\to 0$) behaviors 
\begin{equation}
\delta g(\ell)\aseq\ell^\omega \delta g
\end{equation}
and
\begin{equation}\label{eq:lambdasf}
\lambda_j(\ell)\aseq
\ell\begin{cases}
\check{\lambda}_j+a_{j,g}\,\delta g\,
\frac{\ell^{\omega-1}-1}{\omega-1}&\text{for }\omega\ne 1,\\[\medskipamount]
\check{\lambda}_j+a_{j,g}\,\delta g\,\ln\ell,&\text{for }\omega=1,
\end{cases}
\end{equation}
respectively. Thickness-dependent finite-size quantities such as the excess free energy~\eqref{eq:fexdef}
at bulk criticality are expected to have corrections to scaling linear in $\check{\lambda}[1/\mu L]$. According to Eq.~\eqref{eq:lambdasf}, they  become anomalous at $d=3$, involving  $L^{-1}\ln(\mu L)$ contributions.

This concludes the general part of our discussion of corrections to scaling. We next turn to the numerical determination of the large-$n$ solutions. 

\section{Partially discretized model}\label{sec:pdmod}

In order to determine the solutions of the large-$n$ equations \eqref{eq:ev} and \eqref{eq:V-tau} as well as the excess free energy density \eqref{eq:fexdef} by numerical means, a discretization of the model~\eqref{eq:Ham} is needed. Here we describe our computations based on the first of our discretization schemes where only the $z$~coordinate is discretized and the $\bm{p}$~integrals are dimensionally regularized. To distinguish the so-defined discretized version of our model~\eqref{eq:Ham} from the one obtained by means of a lattice discretization, we shall refer to the former and latter as models A and B, respectively. The latter (model B) will be dealt with in Sec.~\ref{sec:lattmod}.

We discretize $z$ in units of a lattice constant $a$. Thus model A consists of $N$ layers located at 
$z \equiv (l-\frac{1}{2}) a$, with $l=1,\dotsc,N\equiv L/a$.
We now need the discrete analog of the Schrödinger operator $-\partial_z^2+V(z)$ in Eq.~\eqref{eq:ev} subject to the boundary conditions~\eqref{eq:bc}. To determine it, let us temporarily consider 
a lattice model of $n$-vector spins $\bm{s}_{\bm{x}}$ interacting via ferromagnetic bonds which we assume to take the values $K_1$, $K_2$, and $K$ (in units of $k_{\mathrm{B}}T$) for all NN bonds in the layers $l=1,N$, and elsewhere, respectively, where  $\bm{x}=(\bm{y},z)$ are  the sites  on a simple cubic lattice $\subset (a\mathbb{Z})^d$. Upon introducing Lagrange multipliers $\lambda_{z}$ for the constraints $\sum_{\bm{y}}\bm{s}_{\bm{y},z}^2/\sum_{\bm{y}}1=n$, we arrive at the Hamiltonian 
 \begin{equation}\label{eq:Hlat}
\mathcal{H}_{\mathrm{lat}}=\frac{1}{2}\sum_{\bm{x},\bm{x}'}(2\lambda_{z}\,\delta_{\bm{x},\bm{x}'}-K_{\bm{x},\bm{x}'})\,\bm{s}_{\bm{x}}\cdot\bm{s}_{\bm{x}'},
\end{equation}
where $K_{\bm{x},\bm{x}'}$ represents the NN bonds and vanishes otherwise. We divide the part of Eq.~\eqref{eq:Hlat} depending on the interaction constants $K$, $K_1$, and $K_2$ into contributions involving $(s_{\bm{x}}-s_{\bm{x}'})^2$ and a site-diagonal remainder. The latter involves the sums of all bonds connected to site $\bm{x}$. For the chosen NN bonds of our model, these sums yield identical results for all sites belonging to the interior layers $l = 2,\dots,N-1$,
 but different ones for the boundary layers $l=1$ and $l=N$. In terms of the dimensionless enhancement parameter (cf.\ \cite{LR75} or \cite[Eq.~(2.18)]{Die86a}) 
\begin{equation}\label{eq:ca}
\cb_ja=1-2(d-1)(K_j/K-1),
\end{equation}
the result becomes
\begin{equation}
\sum_{\bm{x}'}K_{\bm{x},\bm{x}'}/K=2d -a(\delta_{l,1}\,\cb_1+\delta_{l,N}\,\cb_2).
\end{equation}
The contributions involving $(s_{\bm{y},z}-s_{\bm{y},z'})^2$ in adjacent layers $l$ and $l'$ yield the quadratic form $K\sum_{l=1}^{N}(s_{\bm{y},z}-s_{\bm{y},z'})^2$. 

Upon introducing $\bm{\phi}_{z}(\bm{y})\,\rmd^{d-1}(y/a)=K^{1/2}\bm{s}_{\bm{y},z}$, we can now go over to a continuum description with respect to $\bm{y}$. The discrete analog of the Schrödinger operator in Eq.~\eqref{eq:ev} becomes the  $N\times N$ matrix 
\begin{equation}\label{eq:Hmatrix}
\mat{H}=-\mat{D}^2+\mat{V}+(\cb_1a-1)|1\rangle\langle 1|+(\cb_2a-1)|N\rangle\langle N|,
\end{equation}
with the diagonal potential matrix  $\mat{V}=\mathrm{diag}(V_1,\dotsc,V_{N})$ and the tridiagonal matrix
\begin{equation}\label{eq:D2def}
\mat{D}^2=
\begin{pmatrix}
-2&1&&&\\
1&\ddots&\ddots&\\
&\ddots&\ddots&1\\
&&1&-2
\end{pmatrix}.
\end{equation}

To confirm the consistency with the continuum equations~\eqref{eq:ev} and the boundary conditions~\eqref{eq:bc}, let us compute the action of $\mat{H}$ on a state vector with components $\varphi_{l}$. We find
\begin{equation}\label{eq:evcontlim}
\langle l|\mat{H}|\varphi\rangle=
\begin{cases}
(-\rmd_+ +a\cb_1+V_1)\varphi_1,&l=1,\\
( \rmd_- +a\cb_2+V_{N})\varphi_{N},&l=N,\\
(-\rmd_{\text{c}}^2+V_{l})\varphi_l,&1<l<N,
\end{cases}
\end{equation}
where $\rmd_+$, $\rmd_-$, and $\rmd_{\text{c}}^2$ denote the forward, backward,  and second-order central difference operator, respectively, which act as
\begin{equation}
\rmd_\pm\varphi_l=\pm(\varphi_{l\pm1}-\varphi_l),
\quad \rmd_{\text{c}}^2\varphi_l=\varphi_{l+1}-2\varphi_l+\varphi_{l-1}.
\end{equation}
From the exact results~\eqref{eq:BMV} we can infer that $V_{l}$ should vary $\propto (z/a)^{-2}$ on scales $a\lesssim z\ll L,a|\tau a^2|^{-\nu}$. Hence we expect that $a^{-2}\,V_{l}$ approaches a smooth function $V(z)$ in the continuum limit $a\to 0$, as our results below will confirm. With this assumption,  the limit $a\to 0$ of the last line of Eq.~\eqref{eq:evcontlim} yields indeed the Schrödinger equation~\eqref{eq:ev}. The $a\to 0$ limits of the first and last lines give us the boundary conditions. If we assume that $V_1/a\to v_1$ and $V_{N}/a\to v_2$ with $v_1=v_2=0$, we recover the boundary conditions~\eqref{eq:bc} of the continuum theory. Nonvanishing values $v_1$ and $v_2$ could be absorbed by a redefinition of the enhancement variables $\cb_1$ and $\cb_2$.

We now return to the (${d=3}$)-dimensional case. Owing to the absence of a special transition at $T>0$, the choice of the enhancement variables $\cb_j$ should not be crucial. For simplicity, we choose 
\begin{equation}\label{eq:cjchoice}
\cb_1a=\cb_2a=1
\end{equation}
so that the matrix operator~\eqref{eq:Hmatrix} reduces to $\mat{H}=-\mat{D}^2+\mat{V}$. To understand this choice, recall that $1/\cb_j$ has the meaning of an extrapolation length: the linear extrapolation of a function $\varphi(z)$ which satisfies the boundary condition
$\partial_z\ln\varphi|_{z_j}=\cb_j$ at $z=a/2$ and $z=L-a/2$ vanishes at $z=a/2-1/\cb_1$ and $z=L-a/2+1/\cb_2$, respectively.
 For the choice~\eqref{eq:cjchoice}, this vanishing occurs at the fictitious boundary layers $z=-a/2$ and $z=L+a/2$, respectively. Using  $a^{-2}\mat{D}^2$ as a discrete analog of $\partial_z^2$ therefore provides a lattice realization of Dirichlet boundary conditions at the layers $z-a/2=0$ and $z+a/2=L$, a fact which is well known and exploited in the theory of Feynman path integrals (see, e.g., \cite{Kle06}).

Note that the thickness of our discretized system is $L-a$, $L$, or $L+a$ depending on whether we take the first and $N$th layer, the midplanes $z=a/2$ and $z=L-a/2$, or the fictitious boundary layers $l=0$ and $l=N+1$ to bound it. In our numerical analysis in Sec.~\ref{sec:numan} we will account for such potential microscopic thickness changes  $L\to L\pm a$ by the introduction of a properly chosen effective thickness $\Leff$. This will enable us to absorb a substantial part of the corrections to scaling mentioned above.

In the following, we will again set the lattice constant $a$ to unity, unless otherwise explicitly indicated, and hence identify the thickness $L$ with the number of layers $N$. Furthermore, we shift the system by $a/2$ along the $z$ direction, $z \to z+a/2$, so that $l = z$.

We proceed by computing the integral $\int_{\bm{p}}^{(d-1)}$ in Eq.~\eqref{eq:V-dt-d-p}, using dimensional regularization.  The result
\begin{equation}\label{eq:V-dt-d}
\tau-V_{z}=\frac{g}{6}\,A_{d-1}\sum_{\nu=1}^{L}|\varphi_{\nu ,z}|^2\left[\varepsilon_{\nu}^{\frac{d-3}{2}}-\frac{2^{d}\,\Gamma\!\left(\frac{d-2}{2}\right)}{8\sqrt{\pi}\Gamma\!\left(\frac{d-1}{2}\right)}\right],
\end{equation}
is UV finite at $d=3$ and simplifies to 
\begin{equation}\label{eq:V-eps}
\tau-V_{z}=\frac{g}{24\pi}\sum_{\nu=1}^{L}|\varphi_{\nu ,z}|^2\ln\varepsilon_{\nu}=\frac{g}{24\pi}\langle z|\ln\mat{H}|z\rangle,
\end{equation}
where $\varepsilon_\nu$ are the eigenvalues and $\varphi_{\nu,z}\equiv\langle z|\varphi_\nu\rangle$ the components of the associated orthonormalized eigenvectors $|\varphi_\nu\rangle$ of $\mat{H}$.

The calculation of the excess free energy is somewhat lengthier but straightforward (see Appendix~\ref{sec:fex}). Both the bulk and finite-$L$ free energy densities $\fb$ and $f_L$ have poles at $d=3$ with residua independent of and linear in $\tau$. To eliminate these UV singularities, we subtract from $\fb$ and $f_L/L$ the Taylor expansion of $\fb$ to first order in $\tau$,
\begin{equation}
S(\tau,g )=\fb(0,g )+\tau(\partial_{\tau}\fb)(0,g ),
\end{equation}
defining the renormalized free energy densities
\begin{equation}\label{eq:fbrendef}
\fb^{\mathrm{ren}}(\tau,g )=\fb(\tau,g )-S(\tau,g )\end{equation}
and
\begin{equation}\label{eq:fLrendef}
f_L^{\mathrm{ren}}(\tau,g ,L)=f_L(\tau,g ,L)-L\,S(\tau,g ).
\end{equation}

The subtractions cancel in $\fex$. Thus
\begin{equation}\label{eq:fexfexren}
\fex(\tau,g ,L)\equiv\fex^{\mathrm{ren}}(\tau,g ,L)=f_L^{\mathrm{ren}}(\tau,g )-L\fb^{\mathrm{ren}}(\tau,g ).
\end{equation}
The calculation described in Appendix~\ref{sec:fex} yields \footnote{In the following, $\mat A + b \equiv \mat A + b \mat 1$.} 
\begin{equation}\label{eq:fLrenres}
f_L^{\mathrm{ren}}(\tau,g )=\frac{1}{8\pi}\Tr[\mat{H}(1-\ln\mat{H})]-\frac{3}{2g }\sum_{z=1}^{L}(\tau-V_{z})^{2}
\end{equation}
and 
\begin{align}\label{eq:fbrenres}
\fb^{\mathrm{ren}}(\tau,g )&=\frac{1}{8\pi} \sqrt{  r_{\mathrm{b}}(4+  r_{\mathrm{b}})}-\frac{2+  r_{\mathrm{b}}}{4\pi}\arsinh\!\left( \sqrt{  r_{\mathrm{b}}}/2\right)\nonumber\\
 & -\frac{3}{2g }(\tau- r_{\mathrm{b}})^{2},
\end{align}
where $r_{\mathrm{b}}$, the  inverse bulk susceptibility, is given by
\begin{equation}\label{eq:Vb(dt,g)}
 r_{\mathrm{b}}=\begin{cases}
\tau-\frac{g }{12\pi}\arsinh\!\left(\sqrt{ r_{\mathrm{b}}}/2\right) & \text{for }\tau>0,\\
0 &\text{for } \tau\leq 0.
\end{cases}
\end{equation}

As we have seen in Sec.~\ref{sec:corrtoscal}, the partially discretized and dimensionally regularized model considered here (model A) corresponds to the limiting case of a fixed point at $g^*=\infty$. This suggests to consider the limit $g\to\infty$ to gain higher precision in the numerical calculation of scaling functions. Since the potential $V(z)$ must reduce to Bray and Moore's exact scaling result $V_{\infty,\mathrm{c}}^{\mathrm{ord}}(z)$ given in Eq.~\eqref{eq:BMVord} on scales $1\ll z\ll L,|\tau|^{-\nu}$, it is clear that $V(z)$ has a finite and nonzero $g\to\infty$ limit. Directly at the bulk critical point $\tau=0$, the renormalized  free energy density $f_L^{\mathrm{ren}}$ therefore simplifies to the first term of Eq.~\eqref{eq:fLrenres} at $g=g^*$. 

In order to study the temperature dependence of $\fex$ and related quantities, we must make an appropriate $g$-dependent rescaling of the linear scaling field $\tau$ so that it does not vanish at $g^*$. A convenient way of doing this is to absorb $\xi_+(g)=g/24\pi$, the nonuniversal amplitude  of the  bulk correlation length 
\begin{equation}\label{eq:xiamp}
\xi^{(+)}_{\mathrm{b}}=r_{\mathrm{b}}^{-1/2}\aseq \xi_+(g)\,\tau^{-\nu}\quad  \text{ for }T>T_\mathrm{c},
\end{equation}
 in the temperature scaling field by introducing 
 \begin{equation}\label{eq:tdef}
t=24\pi\tau/g.
 \end{equation}

In order that $f_L^{\mathrm{ren}}$ and $f_{\mathrm{b}}^{\mathrm{ren}}$ have finite $g\to\infty$ limits, we subtract the divergent parts $\propto \tau^2$ (which cancel in $\fex$) and define
\begin{subequations}\label{eq:fLfblimits}
\begin{equation}
f^{\mathrm{ren}}_L(t)\equiv\lim_{g\to\infty}\left[f^{\mathrm{ren}}_L(\tau,g)+L\frac{3\tau^2}{2g}\right]_{\tau=gt/24\pi}
\end{equation}
and the associated bulk quantity
\begin{equation}
\fb^{\mathrm{ren}}(t)\equiv\lim_{g\to\infty}\left[f^{\mathrm{ren}}_{\mathrm{b}}(\tau,g)+\frac{3\tau^2}{2g}\right]_{\tau=gt/24\pi}.
\end{equation}
\end{subequations}

We can now safely take the limit $g\to\infty$ in the above equations and explicitly solve Eq.~\eqref{eq:Vb(dt,g)} at $g=\infty$ for $r_{\mathrm{b}}(t)$. The resulting simplified ${g=\infty}$ analogs of Eqs.~\eqref{eq:fexfexren}--\eqref{eq:Vb(dt,g)} become
\begin{subequations}\label{eq:sctfexinf}
\begin{equation}\label{eq:sct}
t=\langle z| \ln\mat{H} |z\rangle,
\end{equation}
\begin{equation}\label{eq:fext}
\fex(t,L)\equiv \fex^{\mathrm{ren}}(t,L)=f_L^{\mathrm{ren}}(t)-L\,\fb^{\mathrm{ren}}(t)
\end{equation}
with
\begin{equation}\label{eq:fLtren}
f_L^{\mathrm{ren}}(t)=\frac{1}{8\pi}\Tr\!\left[\mat{H} \left(1+t-\ln \mat{H} \right)\right]-\frac{t L}{4\pi}\,,
\end{equation}
\begin{equation}\label{eq:fbtren}
\fb^{\mathrm{ren}}(t)=\frac{1}{4\pi}
\begin{cases}
\sinh (t)-t &\text{for } t>0,\\
0 &\text{for } t\le 0.
\end{cases}
\end{equation}
and
\begin{equation}
r_{\mathrm{b}}=\begin{cases}
4\sinh^{2}(t/2) &\text{for } t>0,\\
0 &\text{for } t\le 0.
\end{cases}
\end{equation}
\end{subequations}
In deriving Eqs.~\eqref{eq:fLtren} and \eqref{eq:fbtren} we used the identity $\Tr[\mat{V}-\mat{H}]=\Tr[\mat{D}^2]=-2L$ implied by Eqs.~\eqref{eq:D2def} and \eqref{eq:sct}.

We numerically determined solutions of both sets of equations~\eqref{eq:fexfexren}--\eqref{eq:Vb(dt,g)} and \eqref{eq:sctfexinf}. Before turning in Sec.~\ref{sec:numan} to an exposition of the results, let us first explain how the analysis gets modified if the lattice discretization of model B is used instead.

\section{Lattice model}\label{sec:lattmod}

The discretized version of the soft-spin model~\eqref{eq:Ham}, which we call model B, is defined through the Hamiltonian
\begin{equation} \label{eq:Hl}
\mathcal{H}_{\mathrm{l}}=\sum_{\bm{x}}\bigg[
\frac{1}{2} \sum_{i=1}^d  (\bm{\phi}_{\bm{x}+\bm{e}_i}- \bm{\phi}_{\bm{x}})^2
+\frac{\tb}{2} \phi_{\bm{x}}^2
+\frac{g}{4! n} \phi_{\bm{x}}^4\bigg].
\end{equation}
Here $\bm{x}=(\bm{y},z)\in\mathbb{Z}^d$ with $1\le x_i\le N_i$, $i=1,\dotsc,d$ labels the sites of a finite simple cubic lattice whose lattice constant we set to $a=1$. In accordance with our previous conventions we write $z=x_d$ and $N_d=L$. Each $\bm{\phi}_{\bm{x}}$ is an $n$-vector spin, and $\bm{e}_i$ 
denotes the unit vector along the $x_i$~direction.  Periodic boundary conditions are imposed along all $x_i=y_i$~directions:
\begin{equation}\label{eq:pbc}
\bm{\phi}_{\bm{x}+N_i\bm{e}_i}=\bm{\phi}_{\bm{x}}\quad\text{for } i=1,\dotsc,d-1.
\end{equation}
For simplicity, we do not consider here the possibility that the coefficients of the three interaction terms of the Hamiltonian take different values in the layers $z=1$ and $z=L$. Accordingly, we impose Dirichlet boundary conditions in the adjacent layers $z=0$ and $z=L+1$, requiring
\begin{equation} \label{eq:lDbc}
\bm{\phi}_{\bm{y},z}=\bm{0} \quad\text{ for } z=0 \text{ and }z=L+1.
\end{equation}

Proceeding as in Secs.~\ref{sec:contmod} and \ref{sec:lnl} yields obvious analogs of Eqs.~\eqref{eq:Z_HS} and \eqref{eq:pot}, which involve a lattice field $\bm{\phi}_{\bm{x}}$ and a site-dependent, yet $\bm{y}$-independent extremum
\begin{equation}\label{eq:psixsp}
\rmi\,\psi_{\bm{y},z}\equiv\rmi\,\psi_{\bm{0},z}=V_z-\tb.
\end{equation}

The reduced free energy per unit cross-sectional hyper-area and number of components in the limit $n\to\infty$ becomes
\begin{equation}
f_L=\frac{1}{2A}\sum_{\bm{p}}{\,\Tr}\ln{\big[}\mat{H}+\varepsilon_{d-1}(\bm{p})\big] - \frac{3}{2g}\Tr[(\tb-\mat{V})^2]
\end{equation}
with $A=N_1 \times \dots \times N_{d-1}$ and 
\begin{equation}
 \varepsilon_{d-1}(\bm{p})=4\sum_{i=1}^{d-1}\sin^2\left(\frac{p_i}{2}\right),
\end{equation}
where the components $p_i$ of the $(d{-}1)$-dimensional wave vector $\bm{p}$ are restricted to the discrete values $p_i=2\pi\nu_i/N_i$, $\nu_i=0,1,\dotsc,N_i-1$. 
Further, $\mat{H}=-\mat{D}^2+\mat{V}$ is the previously used matrix operator defined by Eqs.~\eqref{eq:Hmatrix} and \eqref{eq:D2def} with $K_1=K_2=K$. Note that $f_L$ now depends additionally on all finite-size parameters $N_i$, $i=1,\dotsc,d-1$.

The self-consistency equation for $V_z$ implied by the stationarity condition  $\partial f_L[\mat{V}]/\partial V_z=0$ for the functional $f_L[\mat{V}]$ now takes the form 
\begin{equation}\label{eq:Vzeq}
\tb-V_z=-\frac{g}{6A}\sum_{\bm{p}}\sum_{\nu=1}^{L}\frac{\varphi_{\nu,z}\varphi^*_{\nu,z}}{\varepsilon_\nu+\varepsilon_{d-1}(\bm{p})}.
\end{equation}

Variations $\mat{V}\to\mat{V}+\delta\mat{V}$ with $\delta\mat{V}=\mathrm{diag}(\delta V_1,\dotsc,\delta V_L)$ about the solution $\mat{V}$ of this equation imply the linear change 
\begin{equation}
\delta\varepsilon_\nu=\sum_{z=1}^{L}\varphi_{\nu,z}\varphi^*_{\nu,z}\,\delta V_z
\end{equation}
of the eigenvalues. The Hessian form describing the deviation of $f_L[\mat{V}]$ to second order in $\delta\mat{V}$ can be computed in a straightforward fashion. One obtains
\begin{align}
\delta^2f_L[\mat{V};\delta\mat{V}]={}&-\frac{1}{2A}\sum_{\bm{p}}\Tr [\mat{H}+\varepsilon_{d-1}(\bm{p})]^{-1}\delta\mat{V}]^2\nonumber\\
& -\frac{3}{g}\Tr[(\delta\mat{V})^2].
\end{align}
Since it is negative definite, the solution $\mat{V}$ of Eq.~\eqref{eq:Vzeq} corresponds to a maximum. 

We now take the thermodynamic limit $N_1,\dotsc,N_{d-1}\to\infty$. Equation~\eqref{eq:Vzeq} becomes 
\begin{align}
\tb_\mathrm{c}+\tau-V_z&=-\frac{g}{6}\sum_{\nu=1}^{L} W_{d-1}(\varepsilon_\nu)|\varphi_{\nu,z}|^2\nonumber\\
 &=-\frac{g}{6}\,\langle z|W_{d-1}(\mat{H})|z\rangle,
\end{align}
where $W_{d-1}(\lambda)$ denotes a Watson integral defined in Eq.~\eqref{eq:Wddef}.
From the bulk limit $L\to\infty$ of the foregoing equation, or equivalently from Eq.~\eqref{eq:tbc}, we see that the bulk critical value $\tb_\mathrm{c}$ is given by
\begin{equation}\label{eq:tauclat}
\tb_\mathrm{c}=-\frac{g}{6}\,W_d(0).
\end{equation}
The value of the integral on the right-hand side required for our study of the (${d=3}$)-dimensional case is given in Eq.~\eqref{eq:W30}.

Note also that the coefficient of the $\sqrt{\lambda}$ term of the known expansion \cite{Gut10}
\begin{equation}
W_3(\lambda)-W_3(0)=-\frac{1}{4\pi}\,\sqrt{\lambda}+O(\lambda)
\end{equation}
is consistent with Eq.~\eqref{eq:Ibas} since it is $-A_3$. Upon substituting this result into the bulk equation
\begin{equation}\label{eq:bulklateos}
\tau-r_{\mathrm{b}}=-\frac{g}{6}\big[W_d(r_{\mathrm{b}})-W_d(0)\big],\quad\tau\ge 0,
\end{equation}
with $d=3$, one can immediately convince oneself that the results~\eqref{eq:xiamp} for the asymptotic behaviors of $\xi_{\mathrm{b}}$ and $r_{\mathrm{b}}$ as $\tau\to 0+$ carry over to model B.

The free energy $f_L$ can be conveniently written in terms of integrals of Watson functions, namely 
\begin{equation}\label{eq:Uddef}
U_d(\lambda)\equiv\int_0^\pi\frac{\rmd{q}_1}{\pi}\dotsm\int_0^\pi\frac{\rmd{q}_d}{\pi}\ln\bigg[\lambda+4\sum_{i=1}^d\sin^2\left(\frac{q_i}{2}\right)\bigg],
\end{equation}
which satisfy
\begin{equation}\label{eq:Udprime}
U_d'(\lambda)=W_d(\lambda).
\end{equation}
One finds
\begin{align}\label{eq:fLlat}
f_L(\tau,g)&=\frac{1}{2}\sum_{\nu=1}^{L} U_{d-1}(\varepsilon_\nu)-\frac{3}{2g}\sum_{z=1}^{L}(\tb-V_z)^2\nonumber\\
&=\frac{1}{2}\Tr[U_{d-1}(\mat{H})]-\frac{3}{2g}\Tr[(\tb_\mathrm{c}+\tau-\mat{V})^2]
\end{align}
and
\begin{equation}\label{eq:fblat}
\fb(\tau,g)=\frac{1}{2}\,U_d[r_{\mathrm{b}}(\tau)]-\frac{3}{2g}\,[\tb_\mathrm{c}+\tau-r_{\mathrm{b}}(\tau)]^2,
\end{equation}
where $r_{\mathrm{b}}(\tau)$ is the solution to Eq.~\eqref{eq:bulklateos} or zero, depending on whether $\tau>0$ or $\tau\le 0$.

The function $U_2(\lambda)$, which is needed for our analysis of the (${d=3}$)-dimensional case, can be computed from
\begin{equation}\label{eq:W2lambda}
W_2(\lambda)=\frac{2 }{\pi  (\lambda+4)}\,{\ellipK}\bigg(\frac{4}{\lambda+4}\bigg).
\end{equation}
Here  
\begin{align}\label{eq:Klambdadef}
\ellipK(\lambda)&=\int_0^{1}\frac{\rmd x}{\sqrt{(1-x^2)(1-\lambda^2\,x^2)}}\nonumber\\
&=\frac{\pi}{2}\,{}_2F_1\!\left(\frac{1}{2},\frac{1}{2};1;\lambda^2\right)
\end{align} 
is a complete elliptic integral of the first kind, where ${}_pF_q$ denotes the generalized hypergeometric function. Integration of this equation leads to [cf.\ Eq.~(48) of \cite{HGS11} and \cite{Gut10}]
\begin{align}\label{eq:U2ex}
U_2(\lambda)={}&{}\frac{-2}{(\lambda +4)^2}\,
{}_4F_3\bigg[1,1,\frac{3}{2},\frac{3}{2};2,2,2;\Big(\frac{4}{\lambda+4}\Big)^2\bigg]\nonumber\\
& +\ln (\lambda +4).
\end{align}

To harmonize with our analysis of model A, let us introduce renormalized free energy densities $\fb^{\mathrm{ren}}$ and $f_L^{\mathrm{ren}}$ by analogy with Eqs.~\eqref{eq:fbrendef} and \eqref{eq:fLrendef}, even though this would not be necessary since both quantities are now UV finite. The subtraction function becomes
\begin{equation}
S(\tau,g)=
\frac{1}{2}U_d(0)-\frac{3}{2g}\,\tb_\mathrm{c}^2-\frac{ 3\tau}{g}\,\tb_\mathrm{c},
\end{equation}
A straightforward calculation yields the analogs of Eqs.~\eqref{eq:fLrenres} and \eqref{eq:fbrenres}, namely \cite{remU30,Joy01}
\begin{align}\label{eq:fLrenreslat}
f_L^{\mathrm{ren}}(\tau,g)={}&\frac{1}{2}\Tr [U_{d-1}(\mat{H})]-\frac{{L}}{2}\,U_d(0)\nonumber\\&-\frac{3}{2g}\Tr[(\mat{V}-\tau)^2]+\frac{3}{g}\tb_\mathrm{c}\Tr[\mat{V}]
\end{align}
and
\begin{align}\label{eq:tbrenreslat}
\fb^{\mathrm{ren}}(\tau,g)={}&\frac{1}{2}\big[U_d(r_{\mathrm{b}}(\tau))-U_d(0)-r_{\mathrm{b}}(\tau)\,U_d'(0)\big]\nonumber\\& -\frac{3}{2g}[\tau-r_{\mathrm{b}}(\tau)]^2.
\end{align}
 
According to Eq.~\eqref{eq:w3}, the coefficient $w_3$ takes the negative value $w_3^{(c)}$ for our lattice-discretized model B. Therefore, we cannot set $g$ to the special value $6/w^{(c)}_3$ to suppress corrections to scaling. However, we can still consider the limit $g\to\infty$ to look for simplifications of the above self-consistent equations, even though we should expect more corrections to scaling to remain than for model A at $g=\infty$. To this end, we define the $g=\infty$ functions $f_L^{\mathrm{ren}}(t)$ and $\fb^{\mathrm{ren}}(t)$ as in Eq.~\eqref{eq:fLfblimits}. As an analog of the set of equations~\eqref{eq:sctfexinf} we obtain
 \begin{subequations}\label{eq:ginflat}
 \begin{equation}\label{eq:ginflateos}
 -\frac{ t}{4\pi}=\langle z|W_{d-1}(\mat{H})|z\rangle-W_d(0)
 \end{equation}
 \begin{align}\label{eq:ginflatfex}
f_L^{\mathrm{ren}}(t)={}&\frac{1}{2}\Tr[U_{d-1}(\mat{H})-U_d(0)]\nonumber\\ 
&+\frac{1}{2}\left[\frac{t}{4\pi}-W_d(0)\right]\Tr [\mat{V}],
 \end{align}
  \begin{equation}\label{eq:ginffbren}
\fb^{\mathrm{ren}}(t)=\frac{U_d(r_{\mathrm{b}})-U_d(0)}{2}+\frac{r_{\mathrm{b}}}{2}\bigg[\frac{t}{4\pi}-W_d(0)\bigg],
 \end{equation}
 and
 \begin{equation}\label{eq:ginflatbulkeos}
  - \frac{t}{4\pi}=W_d(r_{\mathrm{b}})-W_d(0)\,,\quad t\ge 0.
 \end{equation}
 \end{subequations}

The numerical solutions of the above equations for model B will be discussed and compared with those for model A in the next section.  

\section{Survey of some exactly known properties}\label{sec:exprop}

Before we turn to these numerical results, it will be helpful to collect 
our knowledge of some analytical properties of the scaling functions $\Theta(x)$ and $\vartheta(x)$. In
Appendix~\ref{sec:lowtas}, we use the mapping of our models A and B in the low-temperature limit (described in Appendix~\ref{sec:derivnlsigma}) to gain information about the asymptotic behaviors of the  functions $\Theta(x)$ and $\vartheta(x)$ in the limit $x\to-\infty$. For the (${d=3}$)-dimensional case, we find that the function $\Theta(x)$ should behave as
\begin{equation}\label{eq:Thetaxmininf}
\Theta(x)\mathop{\aseq}\limits_{x\to-\infty}-\frac{\zeta(3)}{16\pi}\left(1-\frac{2\ln|x|+\Thd_1}{x}\right).
\end{equation}
Our perturbative approach used in Appendix~\ref{sec:lowtas} leaves the value of the universal number $\Thd_1$ undetermined; its exact analytical determination is beyond the scope of the present paper. 

The result~\eqref{eq:Thetaxmininf} implies that the associated Casimir-force scaling function
\begin{equation}\label{eq:varthetad3}
\vartheta(x)=2\Theta(x)-x\,\Theta'(x)
\end{equation}
varies asymptotically as 
\begin{equation}\label{eq:varthetaxmininf}
 \vartheta(x)\mathop{\aseq}\limits_{x\to-\infty}-\frac{\zeta(3)}{8\pi}\left(1-\frac{3\ln|x|+3\Thd_1/2-1}{x}\right).\end{equation}
  
Some other interesting analytical results have been obtained recently \cite{DR14} by exploiting consequences  of short-distance expansions (SDE) and boundary-operator expansions (BOE) \cite{DD83a,Die86a,Car90,EKD93,Die97}. 
To explain these results and their consequences, it is necessary to give some background. Recall that a scaling operator $\mathcal{O}(\bm{y},z)$ with scaling dimension $\Delta[\mathcal{O}]$ can be expanded for small distances from the boundary plane ${z=0}$ in terms of boundary scaling operators $\mathcal{O}^{(\mathrm{s})}_j(\bm{y})$ as
\begin{equation}\label{eq:BOE}
\mathcal{O}(\bm{y},z)\mathop{\aseq}_{z\to 0}\sum_j \,
C_{\mathcal{O}\/,j}(z)\,\mathcal{O}^{(\rm{s})}_j(\bm{y}),
\end{equation}
 where $C_{\mathcal{O}\/,j}(z)$ are $c$-number functions. If $\mathcal{O}^{(\rm{s})}_j$ has scaling dimension $\Delta^{(\mathrm{s})}_j$, then $C_{\mathcal{O}\/,j}(z)$ must scale  $\asprop z^{\Delta^{(\mathrm{s})}_j-\Delta[\mathcal{O}]}$. The potential $V(z)$ corresponds to the expectation value of the energy-density operator. Hence, the BOE can be applied to it. There is convincing evidence that the leading boundary operators $\mathcal{O}_j^{(\mathrm{s})}$ contributing to the BOE of the energy-density operator $\varepsilon(\bm{y},z)= \phi^2$ at the ordinary transition are the unity operator $\openone$ and the $zz$ component $T_{zz}$ of the stress-energy tensor. The contribution from the former yields the critical potential $V^{\text{ord}}_{\infty,\mathrm{c}}(z)$ given in Eq.~\eqref{eq:BMVord}. Away from $T_{\mathrm{c}}$, the corresponding $c$-number function has temperature-dependent corrections. Since this function is a short-distance property, it is expected to be analytic in $t$. The  stress tensor $T_{zz}$, on the other hand, scales $\asprop z^d$ with its engineering dimension $d$, and hence yields a leading thermal singularity $\asprop t^{d\nu}$ \cite{Die97,BD94,DD81c,EKD93}. The upshot is that the self-consistent potential $V(z|t,L)$ for $L=\infty$ should behave as 
 \begin{align} \label{eq:tLcorrV}
\frac{V(z|t,\infty)}{V^{\text{ord}}_{\infty,\mathrm{c}}(z)}\aseq{}& 1+a_1(d)\,t\, (z/\xi_+)^{1/\nu}+O(t^2)\nonumber\\
&+b_0(d)\,(z/\xi_+)^{d}\,t^{d\nu}+\dots
\end{align}
on long scales, where the ellipsis represents terms $\asprop O(t)t^{d\nu}$ and corrections due to other boundary operators. 

Upon including the term $\propto a_1(d)$ in the ansatz for $V(z|t,\infty)$, one can determine $a_1(d)$ from the self-consistency equations. The result for $a_1(3)$ found in \cite{DR14}  is
\begin{equation}
a_1(3)=-\frac{16}{\pi^2}.
\end{equation}

It turns out that the coefficient $a_1(3)$ agrees up to a factor with the amplitude of the leading thermal singularity $\sim t^2 \ln|t|$ of the surface free energy $f_{\mathrm{s}}$. This logarithmic anomaly of $f_{\mathrm{s}}$ arises by a familiar mechanism \cite{CK86} from the interference of the regular contribution $f^{(\mathrm{s})}_2 t^2$  with the singular one $A_\pm^{(\mathrm{s})} |t|^{\nu(d-1)}\asprop t^{2+O(d-3)}$, where the subscripts $\pm$  as usual  indicate that the critical point is approached from positive or negative values of $t$. The $d$-dependent amplitudes of both terms have  pole terms $\propto (d-3)^{-1}$, which cancel to produce a finite $t^2\ln|t|$ singularity at $d=3$. Noting that the contribution $\propto a_1(3)$ in Eq.~\eqref{eq:tLcorrV} appears in the integral $\int_0^\infty \rmd{z}\dots$ giving the excess energy density $\partial f_{\mathrm{s}}/\partial \tau$, one sees that the above residues are proportional to $a_1(3)$ and can determine the proportionality constants \cite{DR14}. One finds that the sum of the leading singular contribution and the regular one have the limit 
\begin{align}\label{eq:logsingfs}
\lim_{d\to 3}\left[A^{(\rm{s})}_\pm(d)|t|^{\frac{d-1}{d-2}}+f^{(\rm{s})}_2(d)\,t^2\right]\nonumber\\
=t^2\left[A_{0,\pm}^{(\rm{s})}+\frac{a_1(3)}{64\pi}\ln |t|\right].
\end{align}

Note that the amplitudes $A^{(\mathrm{s})}_{0,\pm}$ are nonuniversal. However, their difference 
\begin{equation}
\Delta A^{(\rm{s})}_0=A_{0,+}^{(\rm{s})}-A_{0,-}^{(\rm{s})}
\end{equation}
is  given by the $O[(d-3)^0]$ term of the universal ratio $A^{(\rm{s})}_+(d)/A^{(\rm{s})}_-(d)$ and hence universal. To determine  $\Delta A^{(\rm{s})}_0$  exactly, one must go beyond the analysis of \cite{DR14}. As will be shown elsewhere \cite{RD14}, this can be achieved by using  inverse-scattering-theory methods \cite{CS89}. One finds
\begin{align}\label{eq:DelA0ex}
\Delta A^{(\rm{s})}_0&=\frac{1}{16 \pi}\,\int_0^\infty \rmd u \, \frac{\coth u- u^{-1}}{u^2+(\pi/2)^2} \\
& = 0.00944132\ldots\,.\nonumber
\end{align}

From the above results interesting properties of the scaling functions $\Theta(x)$ and $\vartheta(x)$ follow as a consequence of analyticity requirements. To see this, note that the system does not have a phase transition for finite thickness $L$. Hence, the free energy density must be regular at $t=0$ when $L<\infty$. Thus both the thermal singularity of the bulk contribution $Lf_L$ and the thermal singularity of the contribution $2f_{\mathrm{s}}$ to $f_L$ must get canceled by corresponding ones contained in $L^{-2}\,\Theta(tL)$. This idea can be exploited in a straightforward fashion \cite{DR14} to conclude that the function $\Theta(x)$ must behave as  
\begin{align}\label{eq:Thetasing}
\Theta(x) ={}& \D{} + \sum_{k>0} \alpha_k x^k +\frac{x^2\ln|x|}{2\pi^3} \nonumber \\
{}& -2x^2\,H(x)\left(\Delta A_0^{(\rm{s})}+\frac{x}{48\pi}\right),
\end{align}
where $H(x)$ denotes the Heaviside step function. Substituting this result into Eq.~\eqref{eq:relvarthetaTheta} with $d=3$, one finds that the second derivative $\vartheta''(0)$ of the associated Casimir force scaling function $\vartheta(x)$, Eq.~\eqref{eq:varthetad3}, takes the universal value 
\begin{equation}\label{eq:thetapp(0)}
\vartheta''(0)=-\frac{1}{\pi^3}.
\end{equation}

The BOE used above can also be applied to $V(z|0,L)$. It gives
\begin{equation}
\frac{V(z|0,L)}{V^{\text{ord}}_{\infty,\mathrm{c}}(z)}\aseq 1+\mathcal{B}(d)\,(z/L)^d.
\end{equation}
Here the term $\asprop \mathcal{B}(d)$ describes the effect of the far boundary plane $z=L$ on $V(z)$ near the $z=0$ plane. The coefficient $\mathcal{B}(d)$, called distant-wall correction amplitude, is proportional to $\langle T_{zz}\rangle_{t=0,L}/n=(d-1)\Delta_{\mathrm{C}}$ \cite{Car90b,remT}. From  \cite{Car90b} it is known that the ratio $\mathcal{B}(d)/\Delta_{\rm{C}}$ agrees (up to known factors) with the SDE coefficient $\mathcal{B}_\epsilon^T$ of the energy-density operator $\varepsilon(\bm{y},z)$ associated with $T_{zz}$. According to \cite{DR14}, this coefficient can be gleaned  \cite{MO95} to determine the ratio $\mathcal{B}(d)/\Delta_{\rm{C}}$. The resulting value of the distant-wall amplitude for $d=3$ is 
\begin{equation} \label{eq:dwamp}
\mathcal{B}(3)=-\frac{1024}{\pi}\,\Delta_{\rm C}.
\end{equation} 

The above results~\eqref{eq:varthetaxmininf}, \eqref{eq:thetapp(0)}, and \eqref{eq:dwamp} will be checked and confirmed by our numerical results below.

\section{Numerical analysis}\label{sec:numan}

\subsection{Model A at $\tau=0$}

\begin{figure}
\begin{centering}
\includegraphics[scale=0.6]{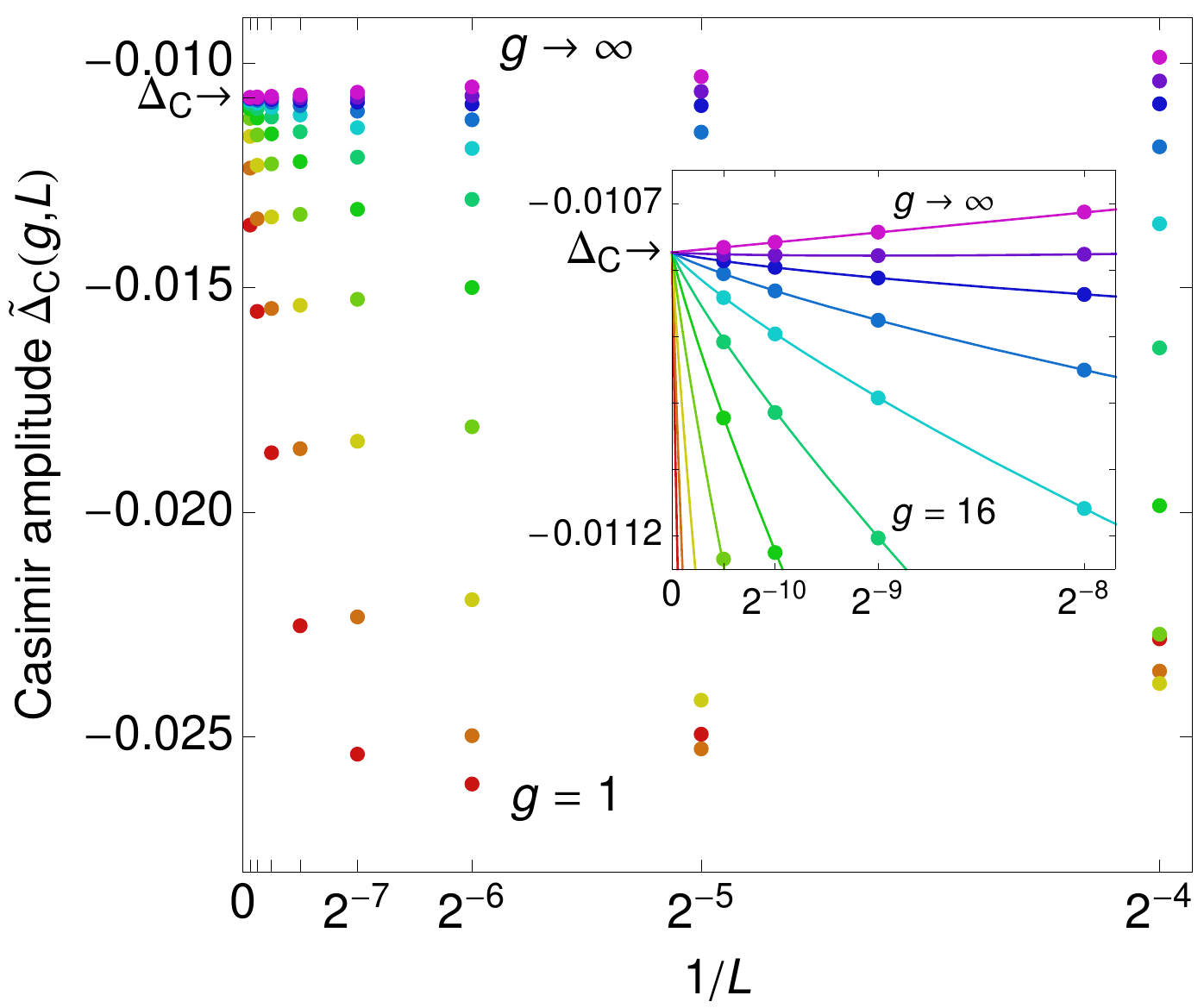}
\par\end{centering}
\caption{\label{fig:Delta(g,L)}(Color online) Effective amplitude $\tilde{\Delta}_{\mathrm{C}}(g,L)$ for different $g=1,2,4,\dots,512,\infty$. Note the strong corrections present at small $g$.}
\end{figure}
In the numerical analysis of model A we first focus on the critical point $\tau=0$, where Eqs.~\eqref{eq:V-eps}-\eqref{eq:fexfexren} simplify to
\begin{subequations}\label{eq:fexV(0,g)}
\begin{align}
\fex(\text{0},g,L) & = \frac{1}{8\pi}\Tr\!\left[\mat{H}\left(1-\ln\mat{H}\right)\right]-\frac{3}{2g}\Tr\!\left[\mat V^{2}\right],\label{eq:fex(0,g)} \\
V_{z} & = -\frac{g}{24\pi}\langle z|\ln\mat{H}\,|z\rangle.\label{eq:V(0,g)}
\end{align}
\end{subequations}
We solved the self-consistency equation Eq.~\eqref{eq:V(0,g)} numerically for different values of $g$ and $L$. From the corresponding results for $\fex(0,g,L)$, Eq.~\eqref{eq:fex(0,g)}, we derived a first estimate for the Casimir amplitude $\D{}$, Eq.~\eqref{eq:DeltaCdef}, using 
\begin{equation}\label{eq:Delta(g,L)}
\tilde{\Delta}_{\mathrm{C}}(g,L)=L^{2}\left[\fex(0,g,L)-\fs(0,g)\right],
\end{equation}
where the surface contribution, Eq.~\eqref{eq:fexfs}, was determined graphically for simplicity.
The results are shown in Fig.~\ref{fig:Delta(g,L)}. Obviously, the convergence  is very unsatisfactory for the case $g=1$ (red circles). The corresponding results for $L\lesssim 100$ seem to approach an incorrect value of approximately $-0.0266$. Only for large $L\gg 100$ the effective Casimir amplitude approaches the correct limit \footnote{Note that in \cite{CHG09}, which
only considered the case $g=1$, the less accurate result $\D{}=-0.012(2)$ was obtained.} $\tilde{\Delta}_{\mathrm{C}}(1,\infty)=-0.0108(1)$. 

\begin{figure}[t]
\begin{centering}
\includegraphics[scale=0.6]{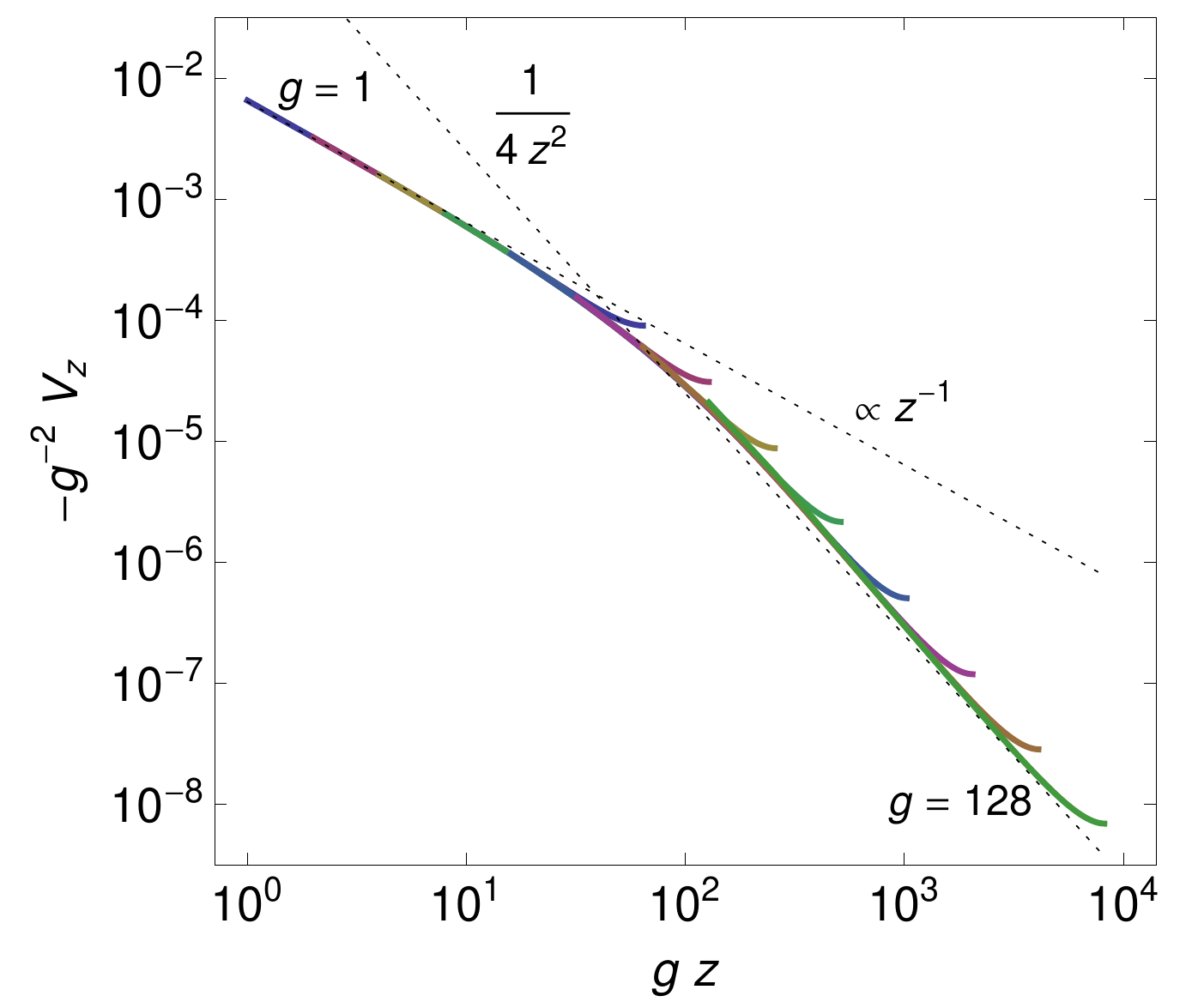}
\par\end{centering}
\caption{\label{fig:V(z)}(Color online) Crossover in $V_{z}$ at criticality
for $L=128$ and different values of $g=1,2,4,\dots,128$.}
\end{figure}
The convergence is much better for $g\gg1$. Those curves in Fig.~\ref{fig:Delta(g,L)} that pertain to the results for $g=64,\dots,512$ show a considerably smoother approach to the limit $L\to\infty$, and do not pass through a minimum. The nonmonotonic or monotonic $L$ dependence of $\tilde{\Delta}_{\mathrm{C}}(g,L)$  when $g$ is small or  large corresponds to a crossover in $V_{z}$. This is illustrated in Fig.~\ref{fig:V(z)}, where we depict the crossover scaling function $V^{\times}$ of $V_{z}$, fulfilling 
\begin{equation}
V_{z}\aseq g^{2}V^{\times}(gz).\label{eq:Vcross}
\end{equation}
The $V_{z}$ curves start out for small $gz$ with a slope of minus one and then bend over to a slope representing  the correct asymptotic behavior $V_{z}\sim-1/4z^{2}$ \cite{BM77a}. This happens at the crossover point $z^{\times}\etwa 40/g$, which is the intersection of the two dotted asymptotes. Since the considered system is symmetric about $z=(L+1)/2$, we expect the crossover to occur at $L^{\times}\sim2z^{\times}\etwa 80/g$, which is indeed the position of the minimum of $\tilde{\Delta}_{\mathrm{C}}(1,L)$ in Fig.~\ref{fig:Delta(g,L)}.

\begin{table*}[t]
{\renewcommand{\arraystretch}{1.2}
\begin{tabular}{@{\hspace{2ex}}c@{\hspace{2ex}}|@{\hspace{2ex}}l@{\hspace{2ex}}l@{\hspace{2ex}}|@{\hspace{2ex}}l@{\hspace{2ex}}l@{\hspace{2ex}}}
\hline\hline
    & $g=32$         &        & $g\to\infty$ & \tabularnewline
$L$ & $\fex(0,32,L)$     & $\D{}$ 
    & $\fex(0,\infty,L)$ & $\D{}$ 
\tabularnewline[0.5ex]
\hline
$2^{2}$ 	& 0.03398692308 	& 				& 0.0434426464161452635463 	& \tabularnewline
$2^{3}$ 	& 0.03473050738 	& 				& 0.0437917553127071125807 	& \tabularnewline
$2^{4}$ 	& 0.03494050671 	& 				& 0.0438954577901547944617 	& \tabularnewline
$2^{5}$ 	& 0.03499017981 	& 				& 0.0439239629614105308154 	& \tabularnewline
$2^{6}$ 	& 0.03500115237 	& 				& 0.0439314545835357953778 	& \tabularnewline
$2^{7}$ 	& 0.03500359692 	& $-0.011062$ 	& 0.0439333762480393760327 	& $-0.01077336957148$\tabularnewline
$2^{8}$ 	& 0.03500415930 	& $-0.010913$ 	& 0.0439338629673000452260 	& $-0.01077340534297$\tabularnewline
$2^{9}$ 	& 0.03500429253 	& $-0.010842$ 	& 0.0439339854485286044466 	& $-0.01077340679713$\tabularnewline
$2^{10}$ 	& 0.03500432476 	& $-0.010808$ 	& 0.0439340161698739302592 	& $-0.01077340684854$\tabularnewline
$2^{11}$ 	& 0.03500433267 	& $-0.010791$ 	& 0.0439340238628944026765 	& $-0.01077340685020$\tabularnewline
$\,2^{12}\,$ &  				&  				& 0.0439340257877384528963 	& $-0.01077340685025$\tabularnewline
$\infty$ 	& 0.03500433527(1) & $-0.01077(1)$	& 0.04393402642965613777877(1) 	& $-0.01077340685024782(1)$\tabularnewline[0.5ex]
\hline\hline 
\end{tabular}
}
\caption{$\fex(0,g,L)$ and estimates of $\D{}$ for $g=32$ (left) and $g\to\infty$ (right) using Eq.~\eqref{eq:fex(0,g)-fit}.
Numerical results for a larger set of thicknesses $L$ are given in the supplemental material \cite{supplMat}. 
The results quoted for $L\to\infty$ and $g\to\infty$ were obtained by analyzing this larger set of data and the ansatz~\eqref{eq:fitLeff}. \label{Table:fex_infty}}
\end{table*}
Calculating $\fex(0,g,L)$ for $g=32$ and different $L$ gives the
results listed in Table~\ref{Table:fex_infty}. These values are analyzed
with the ansatz
\begin{equation}
\fex(0,g,L)=f_{\mathrm{s}}(0,g)+\D{}\, L^{-2}+\sum_{k=3}^{m}f_{k}(g)L^{-k}\label{eq:fex(0,g)-fit}
\end{equation}
using $m$ successive values of $\fex(0,g,L)$. In this simplified procedure we 
neglected logarithmic terms of the form $L^{-k}\ln L$.
The resulting estimates of $\D{}$ for $m=6$ are also given in
Table~\ref{Table:fex_infty}. 

We now turn to the case $g\to\infty$, where Eqs.~\eqref{eq:sctfexinf} simplify to
\begin{subequations} \label{eq:fexV(0,inf)}
\begin{align}
\fex(0,L) & = \frac{1}{8\pi}\Tr\!\left[\mat{H}\left(1-\ln\mat{H}\right)\right],\\
0 & = \langle z|\ln\mat{H}\,|z\rangle.
\end{align}
\end{subequations}
Analyzing $\fex(0,\infty,L)$  in the same way as $\fex(0,32,L)$ above, we found a much 
faster convergence of $\D{}$ with increasing $L$, as can been seen from the numbers
reported in the last column of Table~\ref{Table:fex_infty}.
This fact indicates that for $g\to\infty$ logarithmic corrections are absent, as predicted in Sec.~\ref{sec:corrtoscal}.
Motivated by this success, we generated data for a larger set of thicknesses $L$ (see \cite{supplMat}).
These numerical calculations were performed with 33 digits precision, yielding about $30$ significant digits in $\fex$.
To analyze this extended set of data we define the effective thickness
\begin{equation}\label{eq:Leff}
\Leff= L+ \delta L + \sum_{k=1}^{m} b_k L^{-k}  .
\end{equation}
The estimates of $\Delta_\mathrm{C}$ and $\delta L$ are then determined by analyzing $\fex(0,L)$ 
for different $L=1600$, $1800$, $\dots$, $3800$, $4096$ 
with the ansatz 
\begin{equation} \label{eq:fitLeff}
\fex(0,L) = f_s(0) + \Delta_\mathrm{C} \Leff^{-2}.
\end{equation}
Our final results
\begin{subequations}\begin{align}
\D{}&=-0.01077340685024782(1),\label{eq:DeltaC}\\
\delta L&=0.7255032704723(3),\label{eq:deltaL}
\end{align}\end{subequations}
were obtained by using $m=5$ and the largest thicknesses $L$ available.
As a benchmark for the errors, the variations of the estimates resulting from analogous analyses with $m=4$ were used. We could verify that in the limit $g\to\infty$ no logarithmic corrections were present, as predicted in Sec.~\ref{sec:corrtoscal}.

\begin{figure}[b]
\begin{centering}
\includegraphics[scale=0.6]{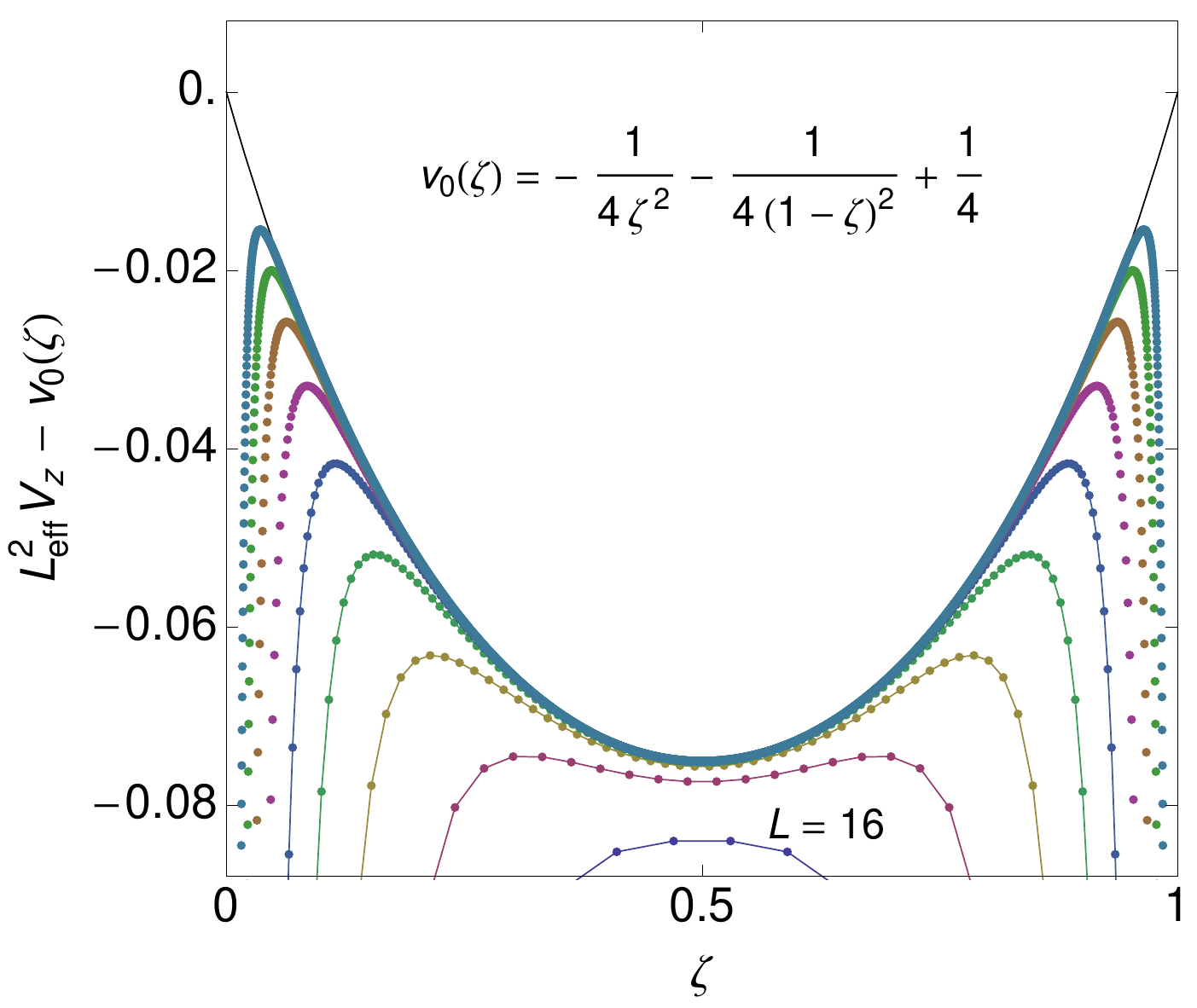}
\par\end{centering}
\caption{\label{fig:v(zeta)}(Color online) Scaled potential $v(\zeta)$, Eq.~\eqref{eq:v(zeta)}, as a function
of $\zeta=z_{\mathrm{eff}}/L_{\mathrm{eff}}$ at criticality, calculated with model A at $g=\infty$
and different values of $L=2^{4},2^{5},\dots,2^{12}$. The dotted line is an extrapolation to $L\to \infty$ (see text).
}
\end{figure}

Finally, we turn to the scaled critical potential 
\begin{equation}\label{eq:v(zeta)}
v(\zeta) = L^2\, V(z) \aseq \Leff^2 V_z, \quad \zeta=\frac{z}{L}
\aseq\frac{z_\mathrm{eff}}{\Leff},
\end{equation}
with $z_\mathrm{eff}=z-1/2+\delta L/2$, which is the numerically exact solution of the continuum model, Eq.~\eqref{eq:V-dt-d-p}.
In Fig.~\ref{fig:v(zeta)} we present an analysis of $v(\zeta)$, written as a sum of the half space contribution
\begin{equation}\label{eq:v0(zeta)}
v_0(\zeta) = -\frac{1}{4\zeta^2} - \frac{1}{4(1-\zeta)^2} + \frac{1}{4} 
\end{equation}
and a power series about the center of the slab,
\begin{subequations}
\begin{equation}\label{eq:v(zeta)fit}
v(\zeta) = v_0(\zeta) + \sum_{k=0}^m \tilde{v}_{2k} \, \left(\zeta - \frac{1}{2} \right)^{2k}
\end{equation}
with coefficients
\begin{align}
\tilde{v}_{0} &= -0.075075422685740932(1), \nonumber\\
\tilde{v}_{2} &=  0.2358287616270474(1), \nonumber\\
\tilde{v}_{4} &=  0.213346985127(1), \nonumber\\
\tilde{v}_{6} &=  0.15090606(1), \nonumber\\
\tilde{v}_{8} &=  0.09356(1), \nonumber\\
\tilde{v}_{10} &= 0.054(1),\nonumber\\
\tilde{v}_{12} &= 0.03(1).
\end{align}
\end{subequations}
In particular, in the center of the film we found $v(1/2) = -7/4+\tilde{v}_0 = -1.825075422685740932(1)$.
These coefficients were determined by first fitting the potential $V_z$ for fixed $L$ using Eq.~\eqref{eq:v(zeta)fit} and then extra\-polating the resulting values to $L\to\infty$. Remarkably, we again find the same value of $\delta L$ as given in Eq.~\eqref{eq:deltaL}.

The consistency of this fit can be checked by comparing it with the exact limiting form
\begin{equation}\label{eq:smallzetav}
v(\zeta) - v_0(\zeta) = \left( \frac{1}{2} + \frac{256  \D{} }{ \pi} \right) \zeta + O(\zeta^2)
\end{equation}
implied by Eq.~\eqref{eq:dwamp}. The fit complies with this predicted asymptotic behavior within the error bars.

\subsection{Model B at $\tau=0$}
Next we analyzed the numerical results that we obtained for 
model B at $\tau=0$. Again, we solved the self-consistency equation iteratively.
Following our discussion in Sec.~\ref{sec:corrtoscal}, we
expect that the leading bulk corrections to scaling do not vanish for 
any value of $g$. However, since they are
minimal at $g=\infty$, we shall focus on this case.

Since leading bulk corrections turned out to be present, we analyzed
our data with ansätze that contain logarithmic corrections. 
For example, in the case of the excess free energy per area, we used
\begin{equation}
\label{DeltaLeff}
\fex(0,g,L) = \fs(0,g) + \Delta_\mathrm{C} \Leff^{-2}
\end{equation}
where the effective thickness of the film is given by
\begin{equation}
\label{LeffX}
\Leff = L + a_0 \ln L+ \delta L + \sum_{k=1}^{m} (a_k \ln L+ b_k) L^{-k}.
\end{equation}
Note that this choice is a  bit ad hoc since for $k > 0$ one might
suppose that the contributions $\propto L^{-k}$ with $k\ge0$ also involve powers $(\ln L)^l$ with $l>1$. However, the
analysis of the data and, in particular, the coincidence of the results
for both models and different values of $g$ justify this choice.

We computed the excess free energy $\fex$ for $L= ...,800$, $900$,
$1000$, $1100$, $1200$, $1300$, $1400$, $1500$, $1600$, $1800$, $2000$,
$2200$, $2500$, $2700$ and $3000$ for $g=\infty$ (see \cite{supplMat}). 
 Analyzing these data we found 
\begin{align}\label{eq:modelBresults}
 \fs &= 0.04757956639699206805522(1), \nonumber \\
 \Delta_\mathrm{C} &= -0.010773406850249(2), \nonumber \\
 a_0  &= -0.123903101(1), \nonumber \\
 \delta  L &= 0.81422072(1).
\end{align}
The numbers were obtained via the ansatz~\eqref{DeltaLeff} with $m=3$.
The error was estimated by comparing with the results obtained for
$m=2$ and $m=4$, and by varying the thicknesses $L$ that are 
included in the analysis.
The result for $\Delta_\mathrm{C}$ is less precise but fully consistent
with the one for model A at $g=\infty$ given in Eq.~\eqref{eq:DeltaC}.	

Next, we analyzed the potential in the middle of the film, obtaining
\begin{align}
v(1/2) &= -1.82507542268(1), \nonumber \\
a_0 &= -0.12390312(1), \nonumber \\
\delta L &= 0.901646(1).
\end{align}
We found that the value of $v(1/2)$ coincides with the one obtained
for model A. The value of $a_0$ is the same as the
one obtained from the analysis of the excess free energy. The two
values of $\delta L$ are similar but definitely not identical.

The analysis of our data for the minimum of the scaling function
$\vartheta(x)$ discussed below corroborate these findings. For 
$x_\mathrm{min}$ as well as $\vartheta(x_\mathrm{min})$ we got values of $a_0$ that
are consistent with those obtained for $x=0$ here, while those 
of $\delta L$ are comparable though not identical.
We conclude that the value of $a_0$ is the same for all quantities
we considered. However, in contrast to model A at $g=\infty$,
$\delta L$ does depend on the quantity that is considered. The 
fact that the values of $\delta L$ do not vary  much might 
be attributed to the fact that the amplitude of the leading bulk correction is small for model B at $g=\infty$. 

\begin{table}
\begin{tabular}{@{\hspace{2ex}}c@{\hspace{2ex}}@{\hspace{2ex}}c@{\hspace{2ex}}@{\hspace{2ex}}l@{\hspace{2ex}}@{\hspace{2ex}}l@{\hspace{2ex}}}
\hline\hline
 model & $g$  & $a_0(g)$        & $a_0^*$ \\
\hline 
A  &     32   & -1.90987(1)     & -10.18597(5)   \\
B  &     60   & -1.1424950(4)   & -10.185918(2)  \\
B  &    240   & -0.37855103(4)  & -10.1859169(5) \\
B  &    600   & -0.22576227(2)  & -10.1859166(4) \\
B  &   1200   & -0.174832685(3) & -10.1859164(1) \\
B  & $\infty$ & -0.123903101(1) & -10.1859163(1) \\ 
\eqref{eq:CorrAmp}&&            & -10.18591635\dots\\
\hline\hline
\end{tabular}
\caption{Results for the amplitude of leading logarithmic 
corrections for both models \label{tableCorrectionRatio}}
\end{table}

\begin{figure*}
\begin{centering}
\hfill
\includegraphics[scale=0.6]{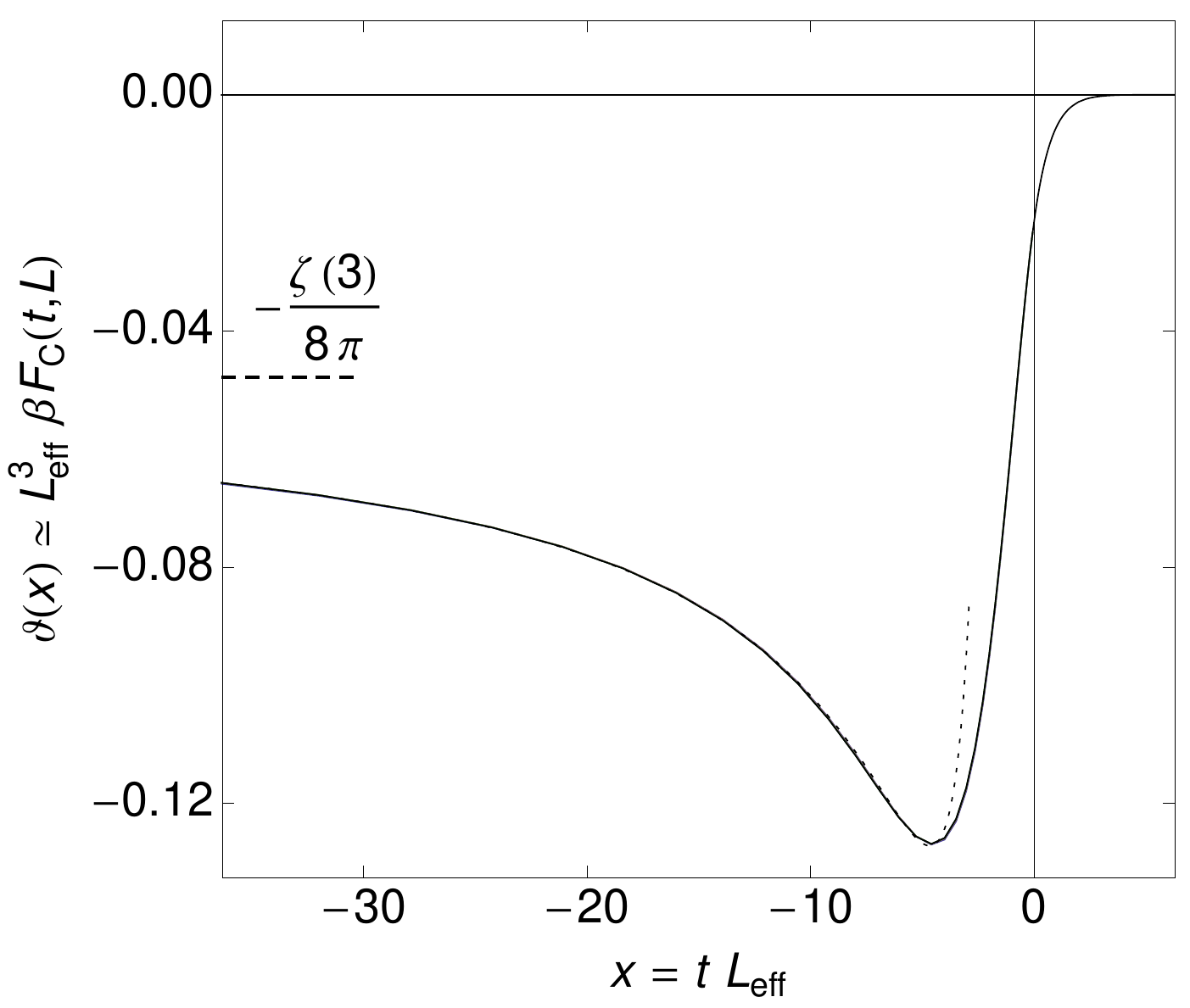}
\hfill
\includegraphics[scale=0.6]{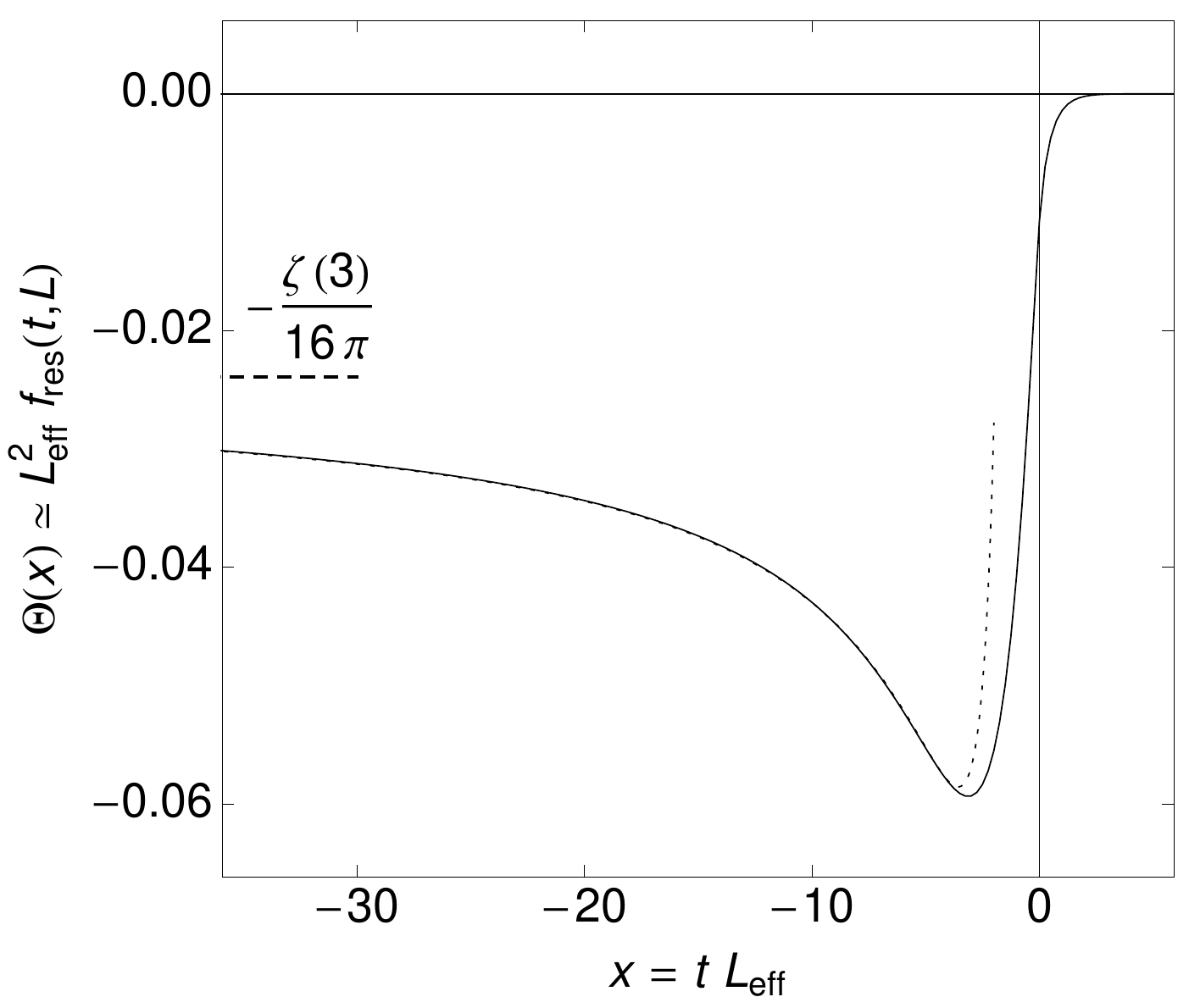}
\hfill{}
\par\end{centering}
\caption{(Color online) Casimir force scaling function $\vartheta(x)$ (left) and residual free energy scaling function $\Theta(x)$ (right) determined from data for $g\to\infty$ and $L=65,97,129,193,257
$ (model A) and $L=
97,129,193,257
$ (model B). The dotted curves represent the asymptotic $x\to-\infty$ forms \eqref{eq:ThetathetaLow}. For further explanations, see main text.
The data for the scaling functions are included in the Supplemental Material \cite{supplMat}.
\label{fig:vartheta(x)} \label{fig:Theta(x)}}
\end{figure*}

Next we studied the dependence of $a_0$ on $g$. Theoretically we expect
that $a_0$ is proportional to the amplitude of the leading bulk
corrections,
\begin{equation}
a_0(g) = a_0^* \left(\frac{6}{g} -w_3 \right),
\end{equation}     
with $w_3$ from Eq.~\eqref{eq:w3}. In order to obtain $a_0(g)$, we analyzed our data for the excess free 
energy generated for various values of $g$. Throughout we got 
consistent, although  less precise, results for the Casimir amplitude
$\Delta_\mathrm{C}$. 
In order to compare also with model A, we reanalyzed the results
obtained for $g=32$, using the ansatz \eqref{LeffX} this time.
Our estimates for $a_0(g)$  are summarized in Table~\ref{tableCorrectionRatio} together with the estimated value of the universal 
corrections to scaling amplitude $a_0^*$.
From the numerics we conjecture the exact value
\begin{equation} \label{eq:CorrAmp}
a_0^* = -\frac{32}{ \pi}
\end{equation}     
for the corrections amplitude.

\section{Results for finite $\tau$}

Figure~\ref{fig:vartheta(x)} shows the  scaling functions $\vartheta(x)$ (left) and $\Theta(x)$ (right) of the Casimir force and the residual free energy that we obtained in the following way from our numerical results for both models A and B.  We first calculated the derivative of the excess free energy with respect to $L$
 according to Eqs.~\eqref{eq:betaFCdef} numerically as
\begin{equation}
\beta \mathcal{F}_\mathrm{C}(t,L)
= -\frac{\fex(t,L+1)-\fex(t,L-1)}{2} + O(L^{-5})
\end{equation}
and then determined the scaling function
\begin{equation}
\vartheta(x)\aseq \Leff^3 \, \beta \mathcal{F}_\mathrm{C}(t,L),
\end{equation}
using $\Leff$ from Eq.~\eqref{eq:Leff} with $m = 0$ for model A, while for model B we took $\Leff$ from Eq.~\eqref{LeffX} with $m=0$, $a_0$ from Eq.~\eqref{eq:modelBresults}, and $\delta L=1$.
This procedure gave the excellent data collapse shown in Fig.~\ref{fig:vartheta(x)} (left).

The curve shows qualitatively the same behavior as for the $XY$ model (corresponding to ${n=2}$) \cite{Huc07,VGMD07,Hasenbusch0905,Has10a}. Using the data from model A, we find a rounded minimum $\vartheta(x_{\mathrm{min}})=-0.1268565841360(1)$ at $x_{\mathrm{min}}=-4.55702477008(1)$, while the curve approaches the Goldstone value $\vartheta(-\infty)=-\zeta(3)/8\pi$ for $x\to-\infty$. 
Note that for the $XY$ model one finds $\vartheta_{n=2}(x_\mathrm{min})\approx-0.65$ at $x_\mathrm{min}\approx -5$ \cite{Huc07,VGMD07,Hasenbusch0905}, where we included a factor $1/n$ in $\vartheta_{n=2}(x)$ because here all energies are defined per spin component; see Eq.~\eqref{eq:fLdef}. 
While the values of $\vartheta(x_{\mathrm{min}})$ differ by a factor of about 5, the locations $x_\mathrm{min}$ for $n=2$ and $n=\infty$ are fairly close.
In the numerical analysis of the minimum within model~A we again found the same value of $\delta L$, Eq.~\eqref{eq:deltaL}, as at criticality.

To compute the scaling function $\Theta(x)$ from $\vartheta(x)$, we used the representation 
\begin{equation}
\Theta(x)=\int_{1}^{\infty}\mathrm{d}s\, s^{-d}\vartheta(xs^{1/\nu}),\label{eq:ThetaRelation}
\end{equation}
which follows upon integration of Eq.~\eqref{eq:fresscf} subject to the condition $\Theta(\infty)=0$.
Note that a direct determination of $\Theta(x)$ from $\fex$ would require the precise calculation of the surface free energy $\fs(t)$ for many values of $t$, a step which is avoided in our approach.
The result is shown in Fig.~\ref{fig:Theta(x)} (right). It looks quite similar
to the Casimir force scaling function $\vartheta(x)/2$ because the
second term in Eq.~\eqref{eq:varthetad3} [involving $\Theta'(x)$] is one order of magnitude
smaller than the first one.

\begin{figure}[t]
\begin{centering}
\includegraphics[scale=0.8333]{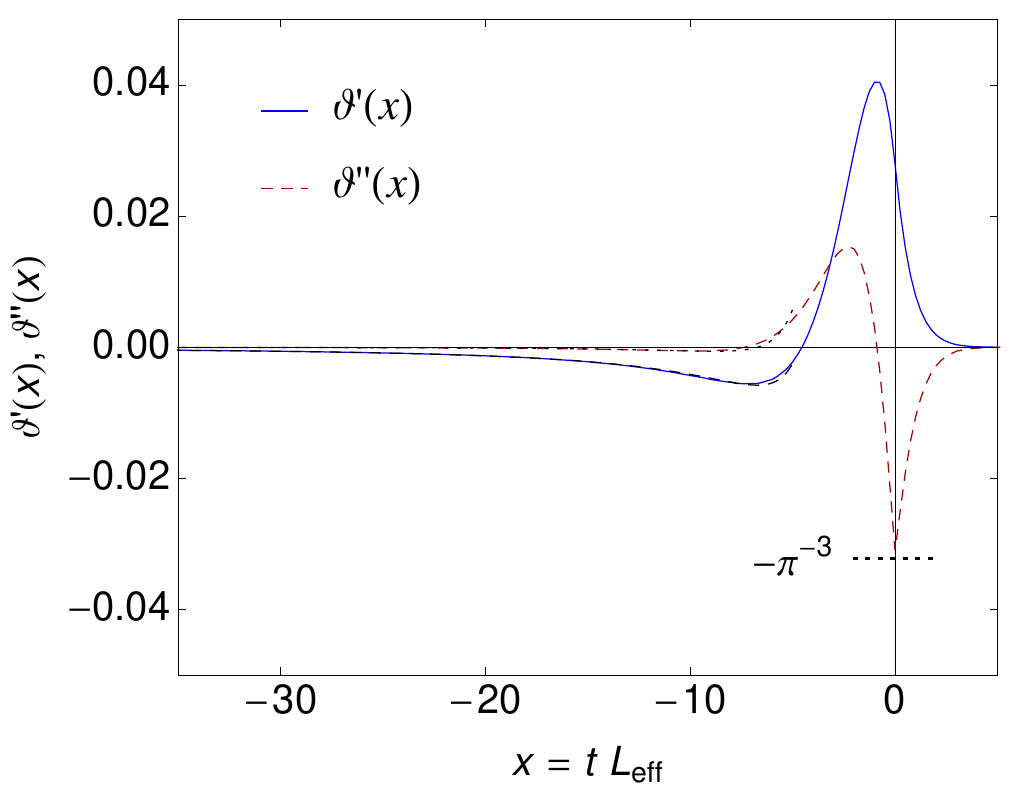} 
\par\end{centering}
\caption{(Color online) First and second derivative $\vartheta'(x)$ (blue solid line) and $\vartheta''(x)$ (red dashed line), determined from data for $g\to\infty$ and $L=257$\label{fig:varthetaPrime(x)}. The horizontal dashed line indicates the exact value $\vartheta''(0)=-\pi^{-3}$ of Eq.~\eqref{eq:thetapp(0)}.}
\end{figure}

In Fig.~\ref{fig:varthetaPrime(x)}, our results for the first and second derivatives of the  Casimir
force scaling function $\vartheta(x)$ are displayed.  To compute $\vartheta'(x)$, we started from  the excess internal energy
\begin{equation}
u_\mathrm{ex}(t,L)\equiv -\frac{\partial \fex(t,L)}{\partial t} = \frac{1}{8\pi}( \Tr[\mat V] - L r_\mathrm{b} ),
\end{equation}
and then used the scaling forms implied by Eqs.~\eqref{eq:fresdef} and \eqref{eq:fresscf} to conclude that $\vartheta'(x)$ can be numerically computed as
\begin{equation}
\vartheta'(x) \aseq -\Leff^2\frac{u_\mathrm{ex}(t, L+1) - u_\mathrm{ex}(t, L-1)}{2}.
\end{equation}

Our considerations based on the nonlinear sigma model (see Appendix~\ref{sec:lowtas}) revealed that the low temperature limits $x\rightarrow-\infty$ of the scaling functions $\Theta(x)$ and $\vartheta(x)$ involve logarithmic anomalies of the form specified in Eqs.~\eqref{eq:Thetaxmininf} and \eqref{eq:varthetaxmininf}, respectively. Guided by these findings, we  analyzed the $x\to-\infty$ limits of our numerical results for $\Theta(x)$ and $\vartheta(x)$ in terms of the ansätze 
\begin{subequations} \label{eq:ThetathetaLow}
\begin{align}
\Theta(x)   \aseq-\frac{\zeta(3)}{16\pi}&\left(1-\sum_{k=1}^m\frac{\Thc_{k}\ln|x|+\Thd_k}{x^k}\right) , \label{eq:ThetaLow}\\
\vartheta(x)\aseq-\frac{\zeta(3)}{8\pi}&\left(1-\frac{3\Thc_{1}\ln|x|+3\Thd_1-\Thc_1}{2x}\right. \nonumber\\
&\quad\left.{}-\frac{4\Thc_{2}\ln|x|+4\Thd_2-\Thc_2}{2x^2}\right). \label{eq:thetaLow}
\end{align}
\end{subequations}

\begin{figure}[t]
\begin{centering}
\includegraphics[scale=0.6]{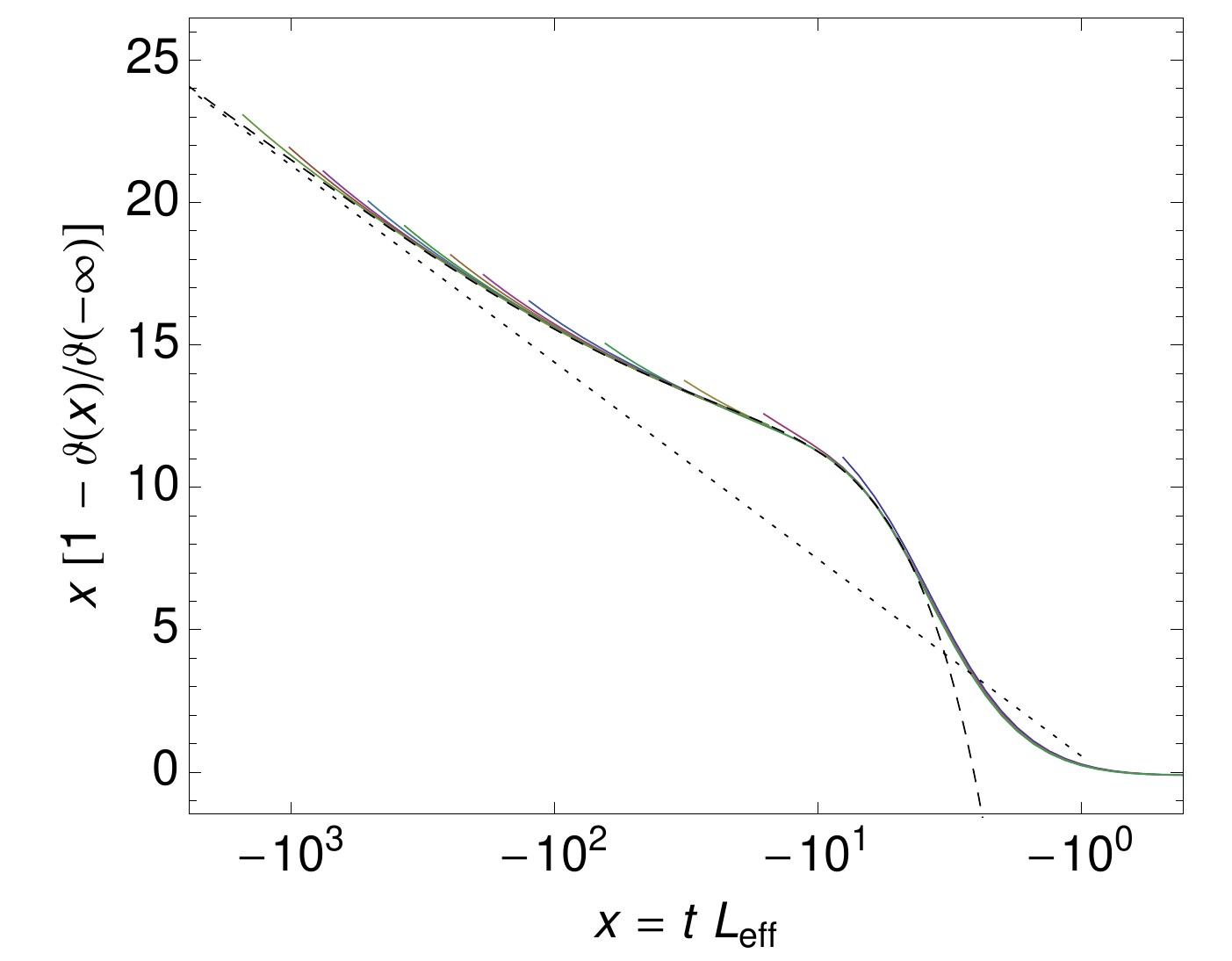}
\par\end{centering}
\caption{(Color online) Asymptotic behavior of $\vartheta(x)$ for $x\to-\infty$. The data are for model A, with $L=9$, 17, 33, 65, 129, 193, 257, 385, 513, 769, 1025, 1537, and are plotted down to $x=-L$. The dashed line is a fit based on Eq.~\eqref{eq:thetaLow} to the data with $m=2$, $\Thc_1=2.0(1)$, $\Thd_1=1.0(1)$, $\Thc_2=-17(2)$, and $\Thd_2=16(2)$. The dotted line with the slope $-3\Thc_1/2=-3$ is a guide to the eyes (see text).}\label{fig:A(x)}
\end{figure}

In Fig.~\ref{fig:A(x)} the quantity
\begin{equation}\label{eq:Upsilon}
\Upsilon(x) = x\left[1-\frac{\vartheta(x)}{\vartheta(-\infty)}\right]
\end{equation}
is shown, 
which becomes a straight line with slope $-3\Thc_1/2$ in the limit $x\rightarrow-\infty$ when plotted versus $-\ln|x|$.
The data for various system sizes $L$ lie on the asymptote down to $x \etwa -L$ and then bend off to larger values. 
Hence large values of $L$ are required to determine the correct asymptotic form
and it is not possible to get the correct low-temperature scaling behavior by an expansion about $T=0$ at constant $L$ as has been done in Ref.~\cite{DBR14} (for details, see \cite{DGHHRS14comm}).

From these results we deduce the parameters $\Thc_1=2.0(1)$ and $\Thd_1=1.0(1)$ \footnote{Note that in \cite{DGHHRS12} incorrect values $\Thc_1\etwa 1.1$ and $\Thd_1\etwa 5.5$ were given, as only data for $x\gtrsim -25$ were available, see Fig.~\ref{fig:A(x)}.}. The former is in accordance with Eq.~\eqref{eq:varthetaxmininf}. The determination of the exact analytical value of the latter is beyond the scope of the present  paper and will be left to a forthcoming paper \cite{RD14}.

Note that for low temperatures the smallest eigenvalue becomes exponentially small, $L^2\varepsilon_1\asprop |x|\exp(x)$. Therefore, its  direct numerical  determination becomes impossible for $x\lesssim-30$. However, one can bypass this problem because its logarithm can be expressed in terms of the logarithms of all other eigenvalues. To see this, note that Eq.~\eqref{eq:sct} implies the sum rule 
\begin{equation}\label{eq:trlogH}
x=\Tr\ln\mat H=\sum_{\nu=1}^L\ln\varepsilon_\nu,
\end{equation}
which we enforced in a standard manner by means of a Lagrange multiplier. In this way, the given large values of $-x$ could be reached without numerical problems.

Finally, we determined the universal amplitude ratio
\begin{align}
\Delta A^{(\rm{s})}_0 
&= -\frac{1}{4}\lim_{x\to0^+}[\Theta''(x)-\Theta''(-x)]\nonumber\\
&= 0.009441(1)
\end{align}
as well as the universal constant
\begin{equation}
\vartheta''(0) = -0.03225(1),
\end{equation}
which are in good agreement with Eqs.~\eqref{eq:DelA0ex} and \eqref{eq:thetapp(0)}.

\section{Summary and Conclusions}\label{sec:sumconc}

In this paper we presented a detailed analysis of the exact large-$n$ solution of the $O(n)$ $\phi^4$ model on a (${d=3}$)-dimensional strip of width $L$ bounded by free surfaces. Our main aim was to determine the scaling functions $\Theta(x)$ and $\vartheta(x)$ of the residual free energy $f_{\mathrm{res}}(t,L)$ and the fluctuation-induced (Casimir) force $\beta \mathcal{F}_\mathrm{C}(t,L)$ for all temperatures $t$. 

Our motivation to study the exact large-$n$ limit is explained in the Introduction. Multi component vector models whose Hamiltonians have a continuous internal symmetry such as $O(n)$ are notoriously difficult to handle in a three-dimensional strip geometry bounded by free surfaces.  The usual challenges one is faced with when dealing with near-critical behavior of systems in such strip geometries is that bulk, boundary, and finite-size critical behavior must be appropriately dealt with, along with the dimensional crossover of the large-scale behavior in a  $d$-dimensional system to that in an effectively $(d{-}1)$-dimensional system. The additional complication which arises 
at $d=3$ in the continuous-symmetry case is that also the low-temperature behavior crucially  matters since it prevents the presence of long-range order at $T>0$ when $L<\infty$. This combination of challenging and intriguing problems one  encounters at $d=3$ quite generally for any $n\ge 2$ persists in the limit $n\to\infty$. An appealing feature of this limit is that all mentioned difficulties can be successfully tackled in a mathematically controlled fashion by means of a single approach. 

Upon solving the required self-consistency equations numerically, we obtained very accurate results for the scaling functions $\Theta(x)$ and $\vartheta(x)$ shown in Fig.~\ref{fig:Theta(x)}. These exhibit all the qualitative features (p1)--(p5) mentioned in the Introduction and known from experiments on the thinning of ${}^4$He wetting films \cite{GC99,GSGC06} and  Monte Carlo simulations of $XY$ models  \cite{Huc07,VGMD07,Hasenbusch0905}. Furthermore, they nicely agree with the various exactly known properties gathered in Sec.~\ref{sec:exprop}.

For large negative $x$ we find logarithmic scaling behavior as predicted by Eqs.~\eqref{eq:Thetaxmininf}--\eqref{eq:varthetaxmininf}, fulfilling the finite-size scaling hypothesis, Eq.~\eqref{eq:FCscalf}, as the scaling functions are solely dependent on the scaling variable $x$.
This is in contrast to Ref.~\cite{DBR14}, where the existence of $\ln L$ contributions to the scaling functions and a violation of the scaling hypothesis was claimed. We could show that this discrepancy stems from the incorrect data analysis done in Ref.~\cite{DBR14}, as the authors utilized data where the condition $|x|\ll L$ does not hold and nonuniversal corrections to scaling become dominant, leading to deviations from the scaling function as displayed in Fig.~\ref{fig:A(x)} \cite{DGHHRS14comm}.

Future work on near-critical Casimir forces of $O(n)$ models on ${d=3}$~dimensional strips could benefit from the results of this paper in several ways.
First of all, to assess the quality of approximate analytical theories such as \cite{Doh14} and \cite{BBSBH10}, one should apply them to the ${n=\infty}$ case and compare their predictions  with our extremely precise numerical results. The same applies to potential future results obtained via appropriate extensions of the numerical functional renormalization techniques used in \cite{JN13} to investigate critical Casimir forces of $O(n)$ systems in slablike geometries subject to periodic boundary conditions. We also believe that our results might provide useful guidance in the development  of improved analytical approaches to the study of fluctuation-induced forces. As we have seen, the large-$n$ theory succeeds in handling dimensional crossovers properly even in the particularly hard case of $O(n)$ system on ${d=3}$~dimensional strips. Clearly, a crucial factor for this capability is its self-consistent nature. This suggest that it may be reasonable, if not indispensable, to incorporate elements of self-consistency in improved analytic approaches for finite $n$.

\begin{acknowledgments}
During the initial phase of this work, HWD, FMS, DG, and SBR benefited from partial support by Deutsche Forschungsgemeinschaft (DFG) via Grant No.~Di 378/5. Subsequently SBR was supported in part by DFG through Grant No.~Ru 1506/1. The work of MH has been supported by DFG via Grants No.~Ha 3150/2 and No.~Ha 3150/3-1. 
We gratefully acknowledge the support via all these grants.

\end{acknowledgments}

\appendix

\section{\\ Bulk and excess free energies of model A}\label{sec:fex}

The  free energy density $f_L$ involves the dimensionally regularized integral $\int_{\bm{p}}^{(d-1)}\ln(p^2+\varepsilon_\nu)$. To compute it, we insert $1=\nabla_{\bm{p}}\bm{p}/(d-1)$ in the integrand and integrate by parts. We thus arrive at 
\begin{equation}
f_L=-\frac{A_{d-1}}{d-1}\,\sum_{\nu=1}^L\varepsilon_\nu^{(d-1)/2}-\frac{3}{2g}\sum_{z=1}^{L}(\tb_\mathrm{c}+\tau-V_{z})^{2}.
\end{equation}
To derive the bulk free energy density~\eqref{eq:fbdef} from this result, we use $L^{-1}\sum_\nu\xrightarrow[L\to\infty]{}\int_0^\pi\rmd{k}/\pi$ and substitute $\varepsilon_\nu$ and $V_z$  by their respective bulk analogs $\varepsilon_{\mathrm{b}}(k)$ [Eq.~\eqref{eq:epsbk}] and $r_{\mathrm{b}}$. The required $k$~integral is of the form
\begin{equation}\label{eq:IDdef}
I_D(r)\equiv \int_0^\pi\frac{\rmd{k}}{\pi}\,[\varepsilon_{\mathrm{b}}(k)]^{(D-3)/2}
\end{equation}
with $D=d+2$. It can be computed using {\sc Mathematica} \cite{Mathematica9}. One obtains
\begin{equation}\label{eq:Idres}
I_d(r)=(r+4)^{\frac{d-3}{2}} \,
   _2F_1\left(\frac{1}{2},\frac{3-d}{2};1;\frac{4}{r +4}\right),
\end{equation}
where $_{2}F_{1}(a,b;c;z)$ denotes the hypergeometric function.
The resulting bulk free energy density therefore becomes
 \begin{equation}\label{eq:fbbare}
\fb(\tau,g)=-\frac{A_{d-1}}{d-1}\,I_{d+2}[r_{\mathrm{b}}(\tau)] -\frac{3}{2g}[\tb_\mathrm{c}+\tau -r_{\mathrm{b}}(\tau)]^{2},
\end{equation}
where $r_{\mathrm{b}}(\tau)$ means the solution to Eq.~\eqref{eq:Vb(dt,g)}.
An analogous calculation of the integral in  Eq.~\eqref{eq:tbc} shows that  the critical value of $\tb$ is given by
\begin{equation}
\tb_\mathrm{c}=\frac{g}{6}A_{d-1}\,I_d(0).
\end{equation}

Since $A_{d-1}$ has a pole  at $d=3$,
\begin{equation}
A_{d-1}=\frac{1}{2\pi(d-3)}+\frac{\gamma_E-\ln(4\pi)}{4\pi}+O(d-3),
\end{equation}
where $\gamma_E$ is the Euler-Mascheroni constant,
we will need the Taylor expansions of $I_{d+2}(r)$ and $I_d(0)$ to $O(d-3)$. A convenient way to determine the $O(d-3)$ terms is to differentiate the right-hand side of Eq.~\eqref{eq:IDdef} and exchange the differentiation with the integration. One obtains
\begin{equation}
\partial_dI_d(r)|_{d=3}=\arsinh(\sqrt{r}/2),
\end{equation}
which, combined with Eq.~\eqref{eq:Idres}, yields
\begin{equation}
I_d(0)=1+O[(d-3)^2].
\end{equation}
In a similar fashion one shows that
\begin{align}
I_{d+2}(r)={}&2+r+\frac{d-3}{2}\Big\{(2+r)[1+2\arsinh(\sqrt{r}/2)]\nonumber\\
& -\sqrt{r(4+r)}\Big\}+O[(d-3)^2].
\end{align}

The bare bulk free energy~\eqref{eq:fbbare} is not regular at $d=3$. Going over to the renormalized quantity $\fb^{\mathrm{ren}}$ defined by Eq.~\eqref{eq:fbrendef} eliminates its pole terms independent of $\tau$ and linear in $\tau$. A straightforward calculation shows that the limit
\begin{equation}\label{eq:fbren}
\fb^{\mathrm{ren}}(\tau,g)=\lim_{d\to 3}\left[\fb(\tau,g)-\fb(0,g)+A_{d-1}\tau/2\right]
\end{equation}
exists and yields the result given in Eq.~\eqref{eq:fbrenres}.

The pole terms we found in $\fb$ must also appear in $f_L/L$ and will get absorbed by the chosen bulk coun\-ter\-terms. In general, $f_L$ can also have $L$-independent poles, which could be eliminated by additive surface counterterms. This happens indeed if we allow for arbitrary values $K_j$ of  surface bonds. However, for our choice~\eqref{eq:cjchoice} corresponding to Dirichlet boundary conditions on a lattice, such surface UV singularities are absent. Consequently, all UV poles must cancel in the excess free energy $\fex$. To show this we substitute our above results for $f_L$ and $\fb$ into the excess free energy~\eqref{eq:fexdef} and expand in $d-3$. This gives 
\begin{align}
\fex(\tau,g,L) ={}& \frac{1}{2}\,A_{d-1}\bigg[\sum_{z=1}^L(V_z+2)-\sum_{\nu=1}^L\varepsilon_\nu\bigg]\nonumber\\
&{} +\frac{1}{8\pi}\sum_{\nu=1}^L\varepsilon_\nu(1-\ln\varepsilon_\nu)-\frac{3}{2g}\sum_{z=1}^L(\tau-V_z)^2\nonumber\\
&{} -L\fb^{\mathrm{ren}}(\tau,g)+O(d-3).
\end{align}
The sole possible source of pole terms is the term proportional to $A_{d-1}$. However, the term in square brackets vanishes because both sums are equal to $\Tr \mat{H}$. Thus the bare $\fex$ is regular at $d=3$ when expressed in terms of $\tau$ and $g$. It reduces to the result given by Eqs.~\eqref{eq:fexfexren}--\eqref{eq:fbrenres}.

\section{\\ 
Low-temperature limit and nonlinear $\sigma$~model}\label{sec:derivnlsigma}

The purpose of this appendix is to derive from the $n$-vector model~\eqref{eq:Ham} an effective low-temperature model which can be used to gain information about the behavior of the scaling functions $\Theta(x)$ and $\vartheta(x)$ of the residual free energy and the Casimir force in the limit $x\to-\infty$. To this end, we follow an established strategy; see, e.g., \cite{Die80}, \cite{DN86}, and \cite[Sec.~8]{Tsv03}. 

In the low-temperature limit, the dominant fluctuations are those associated with the direction of the order parameter. Fluctuations of the modulus $M(\bm{x})\equiv |\bm{\phi}(\bm{x})|$ of the order parameter get frozen in and less important. We therefore decompose $\bm{\phi}(\bm{x})$ into its modulus and a unit $n$-vector $\bm{s}(\bm{x})$, writing
\begin{equation}\label{eq:phifs}
\bm{\phi}(\bm{x})=M(\bm{x})\,\bm{s}(\bm{x}),\quad [\bm{s}(\bm{x})]^2=1.
\end{equation}
We now wish to perform the radial integrations to obtain an effective Hamiltonian that depends solely on $\bm{s}$. To this end it is useful to introduce the functional measure $\mathcal{D}\mu[M]$, the partition function $\mathcal{Z}_{\text{mod}}$ associated with the modulus and corresponding averages via
\begin{align}
\mathcal{Z}_{\text{mod}}&\equiv \int\mathcal{D}\mu[M]\nonumber\\
&\equiv \prod_{\bm{x}\in\mathfrak{V}}\bigg\{\int_{M(\bm{x})\ge 0}[M(\bm{x})]^{n-1}\,\rmd M(\bm{x})\bigg\}\rme^{-\mathcal{H}[M]}
\end{align}
and 
\begin{equation}
\langle\dots\rangle_{\text{mod}}\equiv\mathcal{Z}_{\text{mod}}^{-1}\int\mathcal{D}\mu[M]\dots\,.
\end{equation}
The partition function $\mathcal{Z}$ of Eq.~\eqref{eq:Z} can now be written as
\begin{equation}
\mathcal{Z}/\mathcal{Z}_{\text{mod}}=\int_{\bm{s}^2=1}\mathcal{D}[\bm{s}]\,\rme^{-\mathcal{H}_{\mathrm{eff}}[\bm{s}]}
\end{equation}
in terms of the effective Hamiltonian $\mathcal{H}_{\text{eff}}[\bm{s}]$ defined by
\begin{equation}
\rme^{-\mathcal{H}_{\mathrm{eff}}[\bm{s}]}=\Big\langle\rme^{-\frac{1}{2}\int_{\mathfrak{V}}\rmd^{d}x[M(\bm{x})]^2[\nabla\bm{s}(\bm{x})]^2}\Big\rangle_{\text{mod}}.
\end{equation}

To evaluate the functional integrals over $M$ required for $\mathcal{Z}_{\text{mod}}$ and $\mathcal{H}_{\mathrm{eff}}[\bm{s}]$ we consider the low-temperature limit $\tb\to-\infty$, $g\to\infty$, with $\tb/g$ fixed, and use perturbation theory. At zero-loop order, we must look for extrema of the integrands. The corresponding necessary condition yields in the case of the second functional integral  the classical equations of motion
\begin{equation}\label{eq:zeroloopM}
\big[-\nabla^2+\tb+(\nabla\bm{s})^2\big]M+\frac{g}{6n}M^3-\frac{n-1}{a^d} M^{-1}=0
\end{equation}
with the boundary conditions
\begin{equation}\label{eq:Mbc}
(\partial_z-\cb_1)M|_{z=0}= (\partial_z+\cb_2)M|_{z=L}=0,
\end{equation}
where the term proportional to $M^{-1}$ results from the measure and $a$ is a discretization length (``lattice constant''). The analogous equations for the  functional integral giving $\mathcal{Z}_{\text{mod}}$ differ from the above only in that the $\bm{s}$-dependent term of Eq.~\eqref{eq:zeroloopM} is absent.

The contribution $\propto M^{-1}$ in Eq.~\eqref{eq:zeroloopM} is subleading in the above-mentioned limit and can be dropped. In the absence of the $(\nabla\bm{s})^2$ term, we must then look for a $\bm{y}$-independent solution $M(z)$ of the equation 
\begin{equation}\label{eq:eqmas}
\left[-\frac{1}{|\tb|}\partial_z^2-1+\frac{g}{6n|\tb|}M^2(z)\right]M(z)=0
\end{equation}
subject to the boundary conditions~\eqref{eq:Mbc}. The prefactor of $\partial_z^2$ gives us a length $\propto |\tb|^{-1/2}$ which tends to zero as $\tb\to-\infty$ and hence becomes much smaller than $L$ in this limit. Thus $M(z)$ must take the bulk value $M_{\mathrm{b}}=\sqrt{6|\tb|n/g}$ outside a boundary region of thickness $\ell_0 \propto |\tb|^{-1/2}$ for any values of $\cb_j\in(0,\infty)$. An easy way to see this is to recall from \cite{LR75} or \cite[Eq.~(2.36)]{Die86a} that the solution for the semi-infinite case reads $M(z)=M_{\mathrm{b}}\tanh[(|\tb|/2)^{1/2}(z+z_0)]$ with $\sinh[(2|\tb|)^{1/2}z_0]=(2|\tb|)^{1/2}/\cb_1$. It follows that the excess surface contribution $\int_0^{\ell_0}[M(z)-M_{\mathrm{b}}]\rmd{z}$ varies as $|\tb|^{-1/2}$ and hence vanishes in the limit $\tb\to-\infty$, $g\to \infty$, with $M_{\mathrm{b}}$ fixed. 

In the presence of the $(\nabla\bm{s})^2$~term, the solution to Eq.~\eqref{eq:zeroloopM} is a functional of $(\nabla\bm{s})^2$. However, by expanding about the $(\nabla\bm{s})^2$~independent solution, one sees that the contributions implied by this term also vanish in the considered $\tb\to\infty$ limit. We thus arrive at a nonlinear $\sigma$~model with Hamiltonian
\begin{equation}\label{eq:Hnlsm}
\mathcal{H}_{\mathrm{eff}}[\bm{s}]=\frac{n\rho_{\mathrm{st}}}{2}\int_{\ell_0}^{L-\ell_0}\rmd{z}\int\rmd^{d-1}y\,(\nabla\bm{s})^2,\quad \bm{s}(\bm{x})^2=1.
\end{equation}
Here $n\rho_{\mathrm{st}}$, the reduced spin stiffness \cite{CL95}, is given by $\rho_{\mathrm{st}}=M_{\mathrm{b}}^2/n=6\tb /g$ according to our derivation. The length $\ell_0$ serves as a cutoff to avoid UV singularities with support on the boundary planes $z=0$ and $z=L$ (see Appendix~\ref{sec:lowtas}).

A nonlinear $\sigma$~model of this kind could also be derived from a classical fixed-length spin model on a lattice by making a continuum approximation. For an $O(n)$ spin model of fixed spin length $M_\mathrm{b}$ on a simple cubic lattice with uniform NN interaction constant $J$ (measured in units of $k_{\mathrm{B}}T$) and lattice constant $a$ one would obtain the approximate result $n\rho_{\mathrm{st}}=J M_{\mathrm{b}}^2 a^{2-d}$. We wish to use this model to determine the behavior on long length scales. As minimal length scale or short-distance cutoff  of the model~\eqref{eq:Hnlsm} we can therefore take a coarse-graining length $\ell_0$ much larger than the lattice constant $a$. As spin-stiffness coefficient $\rho_{\mathrm{st}}$ we should therefore take this quantity on the scale $\ell_0$, i.e., determine it by integrating out all degrees of freedom between $a$ and $\ell_0$. Rather than pursuing such an ambitious goal, we shall take $\rho_{\mathrm{st}}$ as an adjustable phenomenological parameter for which we will make a reasonable choice.

The bulk stiffness coefficient $\rho_{\mathrm{st}}$  can be computed for $\tb<\tb_\mathrm{c}$  exactly in a familiar manner in the limit $n\to\infty$ from the small-momentum behavior of the perpendicular correlation function at $h=0$ \cite{Jos66,Sac97}. One finds \begin{equation}
\rho_{\mathrm{st}}=\frac{6}{g}(\tb_\mathrm{c}-\tb)=-\frac{6}{g}\,\tau,
\end{equation}
which becomes 
\begin{equation}\label{eq:rhost}
\rho_{\mathrm{st}}=-\frac{t}{4\pi}
\end{equation}
when expressed in terms of the temperature variable $t$ introduced in Eq.~\eqref{eq:tdef}. In the low-temperature expansion in inverse powers of $\rho_{\mathrm{st}}$ of the next appendix  we will substitute this result for $\rho_{\mathrm{st}}$. As cutoff $\ell_0$ we shall take the length 
\begin{equation}\label{eq:ell0}
\ell_0=-(c_tt)^{-1}
\end{equation}
where  $c_t\etwa 1$.

\section{\\
Nonlinear $\sigma$ model approach to the low-temperature limits of the scaling functions $\Theta(x)$ and $\vartheta(x)$}\label{sec:lowtas}

In this appendix we will use the nonlinear $\sigma$~model derived in Appendix~\ref{sec:derivnlsigma} to determine the asymptotic behavior of the scaling functions $\Theta(x)$ and $\vartheta(x)$ in the limit $x\to-\infty$.

Supposing that a uniform magnetic field $h$ acts along the $s_n\equiv \sigma$~direction, we make the replacement $(\nabla\bm{s})^2\to (\nabla\bm{s})^2-h\sigma$ in the action~\eqref{eq:Hnlsm} and decompose 
$\bm{s}=(\bm{\pi},\sigma)$ into an $(n-1)$-dimensional transverse component $\bm{\pi}$ and a one-dimensional longitudinal one $\sigma=\sqrt{1-\pi^2}$. We now expand the action in powers of $\bm{\pi}$. From the Gaussian part of the action we can identify the free propagator. It has a mass squared equal to $h$ and is subject to Neumann boundary conditions. Expressed in terms of the bulk propagator $G_{\mathrm{b}}$, it reads
\begin{align}\label{eq:GNNL}
G_{\mathrm{NN}}(\bm{x},\bm{x}'|L)=&\sum_{j=-\infty}^\infty\big[G_{\mathrm{b}}(\bm{x}-\bm{x}'-2jL\bm{e}_z)\nonumber\\
&+G_{\mathrm{b}}(\bm{x}-\bm{x}'+2z'\bm{e}_z-2jL\bm{e}_z)\big].
\end{align}
At $d=3$, the bulk propagator simply becomes
\begin{equation}\label{eq:Gb}
G_{\mathrm{b}}(\bm{x})=\frac{\exp(-|\bm{x}|\sqrt{h})}{4\pi \rho_{\mathrm{st}}|\bm{x}|},\quad d=3.
\end{equation}

To gain information about the asymptotic behavior of the scaling function $\Theta(x)$ in the limit $x\to-\infty$, we now set $d=3$ and $h=0$ and compute the Taylor expansion of $f_{\mathrm{res}}L^2$ to first order in $1/\rho_{\mathrm{st}}$. The zeroth-order term is the known Casimir amplitude $\Delta_{\mathrm{G}}^{\mathrm{NN}}(d=3)=-\zeta(3)/16\pi$ of a Gaussian model subject to Neumann boundary conditions. The term linear in $1/\rho_{\mathrm{st}}$ results from the term $(\nabla\pi^2)^2/8\rho_{\mathrm{st}}$ of the action density in Eq.~\eqref{eq:Hnlsm}. It involves the integral
\begin{equation}
J_{\ell_0}=\frac{1}{8\rho_{\mathrm{st}}}\int_{\ell_0}^{L-\ell_0}\rmd{z}\,[\partial_zG_{\mathrm{NN}}(\bm{x},\bm{x}|L)]^2.
\end{equation}
Using Eqs.~\eqref{eq:GNNL} and \eqref{eq:Gb}, one easily computes
\begin{align}
8\pi L^2\partial_zG_{\mathrm{NN}}(\bm{x},\bm{x}|L)&=\psi'(1-\zeta)-\zeta^{-2}-\psi'(1+\zeta)\nonumber\\ &=f(1-\zeta)-f(\zeta)
\end{align}
with $\zeta=z/L$ and
\begin{equation}
f(\zeta)=\zeta^{-2}+\psi'(1+\zeta),
\end{equation}
where $\psi(\zeta)=\Gamma'(\zeta)/\Gamma(\zeta)$ is the digamma function.
Thus,  $J_{\ell_0}$ can be written as
\begin{equation}
J_{\ell_0}=\frac{2}{8\rho_{\mathrm{st}}}\,\frac{1}{64\pi^2L^3}\int_{\ell_0/L}^{1/2}\rmd{\zeta}\,[f(\zeta)-f(1-\zeta)]^2
\end{equation}

A straightforward calculation yields
\begin{equation}
J_{\ell_0}=\frac{\rho_{\mathrm{st}}^{-1}}{256\pi^2L^3}\bigg[\frac{L^3}{3\ell_0^3}-\frac{8}{3}-8\zeta(3)\ln\frac{1/2}{\ell_0/L}+2K_{\ell_0}+R_{\ell_0}\bigg]
\end{equation}
with
\begin{equation}
K_{\ell_0}=\int_{\ell_0/L}^{1/2}\rmd{\zeta}\,\frac{\psi'(1+\zeta)-\psi'(1-\zeta)-2\psi''(1)\,\zeta}{\zeta^2} 
\end{equation}
and
\begin{equation}
R_{\ell_o}=\int_{\ell_0/L}^{1/2}\rmd{\zeta}\left[\psi'(1+\zeta)-\psi'(1-\zeta)\right]^2.
\end{equation}
Upon subtracting from the result the surface term ${J_{\ell_0}}|_{L=\infty}=[768\pi^2\rho_{\mathrm{st}}\ell_0^3]^{-1}$ along with a logarithm, we can take the limit $\ell_0\to 0$ to obtain
\begin{align}
\lim_{\ell_0\to 0}\left[J_{\ell_0}-J_{\ell_0}\big|_{L=\infty}-\frac{\zeta(3)}{32\pi^2\rho_{\mathrm{st}}L^3}\,\ln\frac{2\ell_0}{L}\right]\nonumber \\ =\frac{1}{256\pi^2 L^3}\,\frac{-r_0}{\rho_{\mathrm{st}}}
\end{align}
with
\begin{equation}\label{eq:r0ex}
r_0=\frac{8}{3}-2K_0-R_0
\end{equation}
It follows that
\begin{align}\label{eq:fresas}
f_{\mathrm{res}}\,L^2={}&-\frac{\zeta(3)}{16\pi}-\frac{1}{L\rho_{\mathrm{st}}}\bigg[\frac{\zeta(3)}{32\pi^2 }\ln\frac{L}{2\ell_0}\nonumber\\& -\frac{r_0}{256\pi^2}+o(L^{0})\bigg] +o(\rho_{\mathrm{st}}^{-2}).
\end{align}

The integrals $K_0$ and $R_0$ can be numerically computed. One obtains
\begin{align}
R_0&=1.7854912528\dots\,,\label{eq:R0app}\\
K_0&=-1.2806128714\dots\,.\label{eq:K0app}
\end{align}
Upon substituting the above numerical results for $K_0$ and $R_0$ into Eq.~\eqref{eq:r0ex}, we arrive at the value
\begin{equation}
r_0=3.4424011568\dots\,.
\end{equation}
{}

We can now substitute Eqs.~\eqref{eq:rhost} and \eqref{eq:ell0} for $\rho_{\mathrm{st}}$ and $\ell_0$. The result
tells us that the asymptotic form of the scaling function $\Theta(x)$ for $x\to -\infty$ does indeed involve a leading logarithmic anomaly of the form specified in Eq.~\eqref{eq:Thetaxmininf}.  Since the coefficient of the subtracted $\ln L$ term in Eq.~\eqref{eq:fresas} is independent of the precise choice of the cutoff length $\ell_0$, i.e., the amplitude $c_t$ in Eq.~\eqref{eq:ell0}, we can trust that our perturbative approach here gives the precise value of the universal coefficient of the $x^{-1}\ln|x|$ term in Eq.~\eqref{eq:Thetaxmininf}. By contrast, the choice of the nonuniversal coefficient $c_t$ affects the amplitude of the contribution $\propto x^{-1}$ in Eq.~\eqref{eq:Thetaxmininf} because of the $\ell_0$ dependence of the term $\ln\frac{L}{2\ell_0}$ in Eq.~\eqref{eq:fresas}. Substituting of Eq.~\eqref{eq:ell0} for $\ell_0$ we arrive at  the $c_t$-dependent value
\begin{equation}
 \Thd_1(c_t) = \frac{r_0}{4\zeta(3)} - 2\ln(c_t/2).
\end{equation}
Thus the perturbative approach used here does not enable us to safely determine the universal value of $d_1$. We can at best hope to get a rough estimate by making plausible choices for $c_t$. Two such estimates are
$\Thd_1(1)\etwa 2.102$ and $\Thd_1(2)\etwa 0.716$. Though not precise, they are not unreasonably far from the value $\Thd_1\etwa 1.0$ our numerical data suggest (cf.\ caption of Fig.~\ref{fig:A(x)}). 

\bibliography{bank,remarks}

\begin{thebibliography}{91}
\bibitem[{rem({\natexlab{b}})}]{remmu}
\bibinfo{note}{We denote the momentum scale as $\mu$. In RG schemes that avoid
  taking the limit $\Lambda\to\infty$ of the large-momentum cutoff $\Lambda$,
  one can, of course, simply take $\Lambda$ instead of $\mu$,}.
\bibitem[{rem({\natexlab{c}})}]{remU30}
\bibinfo{note}{The quantity $U_3(0)$ can be expressed as $U_3(0)={L_1(1)+\ln
  6}$ in terms of the number $L_1(1)=1.67338930297019673228\ldots$ given to 201
  digits in \cite{Joy01}. Its ${d=2}$~analog satisfies the relation
  $U_2(0)=4G/\pi$ where $G= 0.915 965 594\ldots$ is Catalan's constant.}
\bibitem[{rem({\natexlab{d}})}]{remT}
\bibinfo{note}{Note that our definition of the stress-energy tensor differs
  from the one used by Cardy in \cite{Car90b} by a factor
  $-2\pi^d/\Gamma(d/2)$, but agrees with that of \cite{EKD93}.}
\end{thebibliography}
\end{document}